\documentclass[sigconf]{acmart}

\usepackage[toc,page]{appendix}
\usepackage{subfigure}
\usepackage{multirow}
\usepackage{balance}

\setcopyright{none} 

\begin{document}
\title{Profile Matching Across Unstructured Online Social Networks: Threats and Countermeasures}
\titlenote{Institutional Review Board (IRB) approval has been obtained to conduct this work.}

\author{Anisa Halimi}
\affiliation{%
	\institution{Bilkent University}
}
\email{anisa.halimi@bilkent.edu.tr}
\author{Erman Ayday}
\affiliation{%
	\institution{Bilkent University}
}
\email{erman@cs.bilkent.edu.tr}

\begin{abstract}
In this work, we propose a profile matching (or deanonymization) attack for unstructured online social networks (OSNs) in which similarity in graphical structure cannot be used for profile matching. We consider different attributes that are publicly shared by users. Such attributes include both obvious identifiers such as the user name and non-obvious identifiers such as interest similarity or sentiment variation between different posts of a user in different platforms. We study the effect of using different combinations of these attributes to the profile matching in order to show the privacy threat in an extensive way. Our proposed framework mainly relies on machine learning techniques and optimization algorithms. We evaluate the proposed framework on two real-life datasets that are constructed by us. Our results indicate that profiles of the users in different OSNs can be matched with high probability by only using publicly shared attributes and without using the underlying graphical structure of the OSNs. We also propose possible countermeasures to mitigate this threat in the expense of reduction in the accuracy (or utility) of the attributes shared by the users. We formulate the tradeoff between the privacy and profile utility of the users as an optimization problem and show how slight changes in the profiles of the users would reduce the success of the attack. We believe that this work will be a valuable step to build a privacy-preserving tool for users against profile matching attacks between OSNs. 
\end{abstract}


\maketitle

\section{Introduction}
An online social network (OSN) is a platform, in which, individuals share vast amount of information about themselves such as their social and professional life, hobbies, diseases, friends, and opinions. Via OSNs, people also get in touch with other people that share similar interests or that they already know in real-life~\cite{ellison2007social}. With the widespread availability of the Internet, especially via mobile devices, OSNs have been a part of our lives more than ever. Most individuals have multiple OSN profiles for different purposes. Furthermore, each OSN offers different services via different frameworks, leading individuals share different types of information~\cite{debnath2008feature}.

The most common (and basic) types of information that is shared by the individuals in OSNs include screen name, age, gender, location, date of birth, profile photo, and e-mail address~\cite{ellison2007social}. However, such information is usually incomplete and inconsistent across different OSNs. Also, in some OSNs (e.g., Facebook), users mostly reveal their real identities (e.g., to find old friends), while in some OSNs users mainly prefer to remain anonymous (especially in OSNs in which users share sensitive information about themselves, such as health status).

It is trivial to link profiles of individuals across different OSNs in which they share their real identities. However, such profile matching is both nontrivial and sometimes undesired if individuals do not reveal their real identities in some OSNs. While profile matching is useful for online service providers to build complete profiles of individuals (e.g., to provide better personalized advertisement), it also has serious privacy concerns. If an attacker can link anonymous profiles of individuals to their real identities (via their other OSN accounts in which they share their real identity), he can obtain privacy-sensitive information about individuals that is not intended to be linked to their real identities. Such sensitive information can then be used against the individuals for discrimination or blackmailing. Thus, it is very important to quantify and show the risk of such profile matching attacks and provide countermeasures against them.

An OSN can be characterized by (i) its graphical structure (i.e., connections between its users) and (ii) the attributes of its users (i.e., types of information that is shared by its users). The graphical structures of most popular OSNs show strong resemblance to social connections of individuals in real-life (e.g., Facebook). Therefore, it is natural to expect that the graphical structures of such OSNs will be similar to each other as well. Existing work shows that this similarity in graphical structure (along with some background information) can be utilized to link accounts of individuals from different OSNs~\cite{narayan2}. However, without sufficient background information, just using graphical structure for profile matching becomes computationally infeasible.

Furthermore, some OSNs or online platforms either do not have a graphical structure at all (e.g., forums) or their graphical structure does not resemble the real-life connections of the individuals. A good example for the latter is health-related OSNs such as patientslikeme.com. In this OSN, people share sensitive information about themselves such as their health conditions, diseases, diagnosis, and the drugs they use, and hence most users do not share their real identity. Users also follow each  other, especially if they have similar diseases (e.g., to get more information about the alternative treatments of the disease). Thus, the OSN has a graphical structure, but this structure has no similarity to the real-life connections of the users, and hence an attacker cannot use the graphical structure to link the accounts of patientslikeme users to their accounts in other OSNs (in which they share their identifiers). However, this does not mean that users of such OSNs are protected against profile matching (or deanonymization) attacks. In this type of OSNs, an attacker can utilize the attributes of the users across different OSNs for deanonymization.

In this work, we propose a profile matching (or deanonymization) scheme that quantifies and shows the risk of the profile matching attack in unstructured OSNs. We show the threat between an auxiliary OSN (in which users share their real identities) and an anonymous OSN (in which users prefer to make anonymous sharings). The proposed scheme matches user profiles across multiple OSNs by using machine learning and optimization techniques. We mainly focus on two types of attacks (i) targeted attack, in which the attacker selects a set of victims from the auxiliary OSN and wants to determine the profiles of the victims in the anonymous OSN, and (ii) global attack, in which the attacker wants to deanonymize the profiles of all the users that are in the anonymous OSN (assuming they have accounts in the auxiliary OSN). Our results show that by using different machine learning (linear regression and support vector machine) and optimization techniques, individuals' profiles can be matched with up to $70\%$ success (depending on the set of attributes used for the attack). We also study the effect different types of attributes (i.e., obvious identifiers and non-obvious identifiers) to the proposed attack. Furthermore, we propose possible countermeasures against the proposed attack. That is, we show how slight changes in users' attributes (e.g., change in the time of some posts) would reduce the success of the proposed attack, and hence prevent profile matching. Note that the proposed framework can also be used in structured OSNs (along with the structural information).

The main contributions of this work can be summarized as follows:
\begin{itemize}
	\item We develop a profile matching attack across unstructured OSNs by using various publicly shared attributes of the users and show how the privacy risk can be quantified.
	\item We study the effect of different sets of publicly shared attributes to the proposed attack. In particular, we show how obvious identifiers (such as the user name and location) and non-obvious identifiers (such as activity patterns across OSNs, interests, or sentiment) of the users help the attacker.
	\item We propose potential countermeasures against the proposed attack. By formulating the proposed countermeasures as an optimization problem, we show how the users can minimize the profile matching threat while maximizing the utility they get from the OSNs.
	\item We evaluate the proposed attack and the countermeasures on two different real-life datasets that are constructed by us.
\end{itemize}

The rest of the paper is organized as follows. In the next section, we discuss the threat model. In Section~\ref{sec:model}, we detail the proposed framework for profile matching in unstructured OSNs. In Section~\ref{sec:evaluation}, we show the results of the proposed attack by using real data. In Section~\ref{sec:countermeasures}, we propose possible countermeasures against the proposed attack in the expense of reduction in the utility. In Section~\ref{sec:related_work}, we summarize the related work and the main differences of this work from the existing work in the area. Finally, in Section~\ref{sec:conclusion}, we discuss the future work and conclude the paper.

\section{Threat Model}\label{sec:threat}

For simplicity, we consider two OSNs to describe the threat: (i)  $A$, the auxiliary OSN that includes the profiles of individuals with their identifiers, and (ii) $T$, the target OSN that includes anonymous profiles of individuals. In general, the attacker knows the identity of the individuals from OSN $A$ and depending on the type of the attack, he wants to determine the real identities of the user(s) in OSN $T$ by only using the public attributes of the users (i.e., information that is publicly shared by the users). The attacker can be a part (user) of both OSNs and it can collect publicly available data from both OSNs (e.g., via crawling). We assume that the attacker is not an insider in $T$. That is, the attacker cannot use the IP address, access patterns, or sign up information of the victim for profile matching (or deanonymization).

We consider two different attacks (i) targeted attack, and (ii) global attack. In the targeted attack, the attacker wants to deanonymize the anonymous profile of a victim (or a set of victims) in OSN $T$, using the unanonymized profile of the same victim in OSN $A$. In the global attack, the attacker's goal is to deanonymize the anonymous profiles of all individuals in $T$ by using the information in $A$.

\section{Proposed Model}\label{sec:model}

\begin{figure*}[ht]
	\centering
	\includegraphics[scale=0.55]{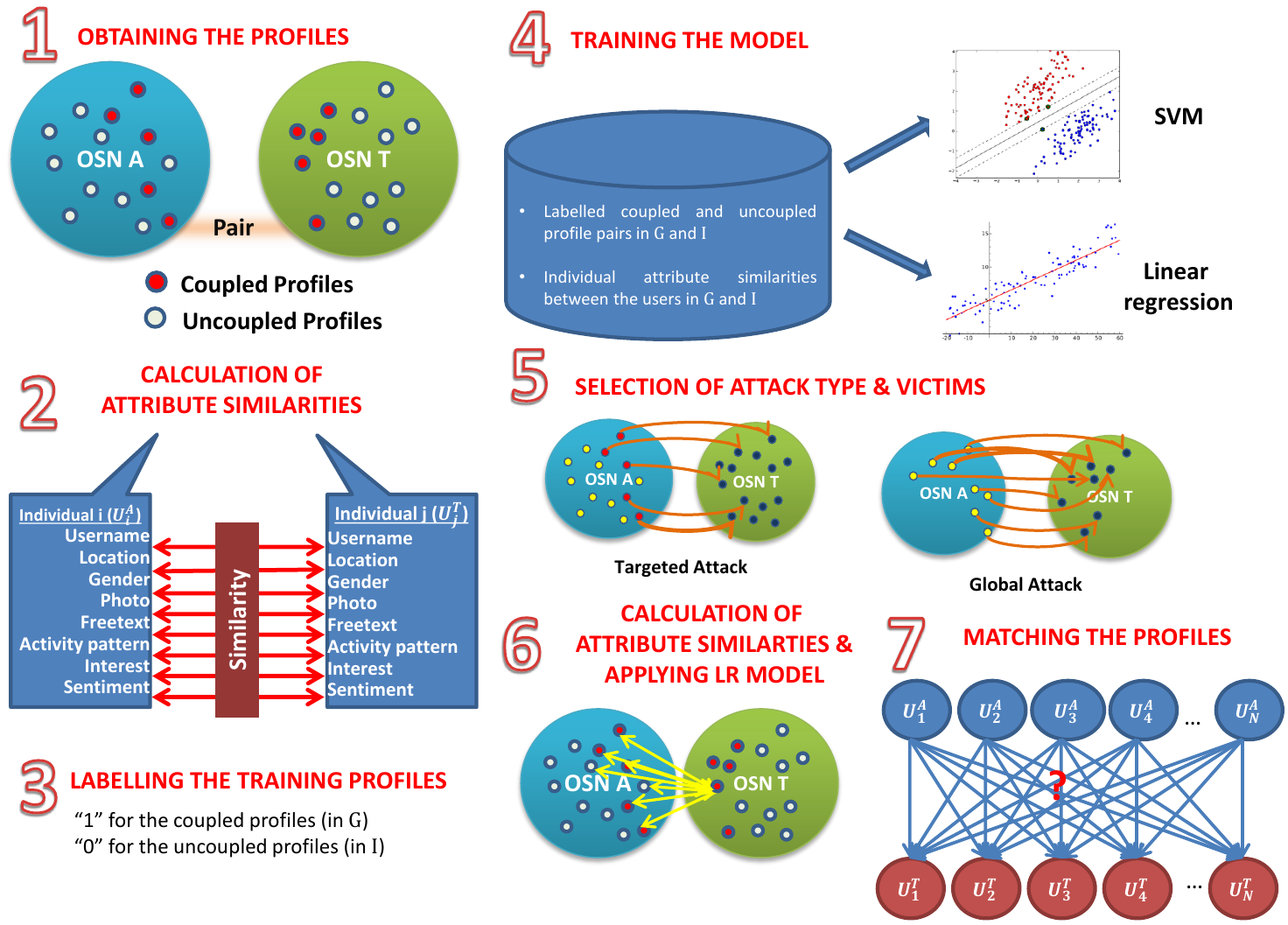}
	\caption{Overview of the proposed profile matching framework.}
	\label{fig:wokmodel}
\end{figure*}

Let $A$ and $T$ represent the auxiliary and the target OSN, respectively, in which people publicly share attributes such as date of birth, gender, and location. Profile of a user $i$ in either $A$ or $T$ is represented as $U_i^k$, where $k\in\{A,T\}$. In this work, we focus on the most common attributes that are shared in many OSNs. Thus, we consider the profile of a user $i$ as $U_i^k=\{n_i^k,\ell_i^k,g_i^k,p_i^k,f_i^k,a_i^k,t_i^k,s_i^k\}$, where $n$ denotes the user name, $\ell$ denotes the location, $g$ denotes the gender, $p$ denotes the profile photo, $f$ denotes the freetext provided by the user in the profile description, $a$ denotes the activity patterns of the user in a given OSN (i.e., time instances at which he is active), $t$ denotes the interests of the user (on that particular OSN), and $s$ denotes the sentiment profile of the user. As discussed, the main goal of the attacker is to link the profiles between two OSNs. General notations that are used in the proposed model are presented in Table~\ref{table:notation}. Furthermore, the overview of the proposed framework is shown in Figure~\ref{fig:wokmodel}.

\begin{table}[ht]
	\centering
	\resizebox{0.40\textwidth}{!}{
		\begin{tabular}{|c|l|}
			\hline
			\multirow{2}{*} {$A$} & Auxiliary OSN, in which profiles of individuals \\ & include  their real identifiers  \\
			\hline
			\multirow{2}{*} {$T$} & Target OSN, in which profiles of individuals \\ & do not include  their real identifiers\\
			\hline
			\( U^A_i \) & Profile of user $i$ in OSN $A$\\
			\hline
			\( U^T_j \) & Profile of user $j$  in OSN $T$\\
			\hline
			\multirow{2}{*} {$\mathrm{G}$} & Set of coupled profiles from $A$ and $T$  \\ & belonging to the same  individual\\
			\hline
			\multirow{2}{*} {$\mathrm{I}$} & Set of uncoupled profiles from $A$ and $T$  \\ & belonging to different individuals\\
			\hline
			\(S( U^A_i, U^T_j)\) & General similarity of $U^A_i$ and $U^T_j$  \\
			\hline
			\(S(n^A_i,n^T_j)\) & Username similarity of $U^A_i$ and $U^T_j$\\
			\hline
			\(S(\ell^A_i,\ell^T_j)\)  & Location similarity of $U^A_i$ and $U^T_j$\\
			\hline
			\(S(g^A_i,g^T_j)\) & Gender similarity of $U^A_i$ and $U^T_j$\\
			\hline
			\(S(p^A_i,p^T_j)\) & Photo similarity of $U^A_i$ and $U^T_j$\\
			\hline
			\(S(f^A_i,f^T_j)\) & Freetext similarity of $U^A_i$ and $U^T_j$\\
			\hline
			\(S(a^A_i,a^T_j)\) & Activity pattern similarity of $U^A_i$ and $U^T_j$\\
			\hline
			\(S(t^A_i,t^T_j)\) & Interest similarity of $U^A_i$ and $U^T_j$\\
			\hline
			\(S(s^A_i,s^T_j)\) & Sentiment similarity of $U^A_i$ and $U^T_j$\\
			\hline
		\end{tabular}
	}
	\caption{Symbols and notations used in this work.}
	\label{table:notation}
\end{table}

In general, the proposed scheme is composed of two main parts: (i) Steps~1-4 (in Figure~\ref{fig:wokmodel}) constitute the training part and they are the offline steps of the algorithm, and (ii) Steps~5-7 are the attack part. In Step~1, profiles and attributes of a set of users are obtained from both OSNs to construct the training dataset. We denote the set of profiles that are extracted for this purpose from OSNs $A$ and $T$ as $\mathrm{A_t}$ and $\mathrm{T_t}$, respectively. We assume that profiles are selected such that some profiles in $\mathrm{A_t}$ and $\mathrm{T_t}$ belong to the same individuals and some do not (more details on collecting such profiles can be found in Section~\ref{sec:datasetcreate}).\footnote{Such profiles are required to construct the ground-truth for training.} We let set $\mathrm{G}$ include pairs of profiles $(U_i^A,U_j^T)$ from $\mathrm{A_t}$ and $\mathrm{T_t}$ that belong to the same individual (i.e., coupled profiles). Similarly, we let set $\mathrm{I}$ include pairs of profiles $(U_i^A,U_j^T)$ from $\mathrm{A_t}$ and $\mathrm{T_t}$ that belong to different individuals (i.e., uncoupled profiles).

In Step~2, for each pair of users in sets $\mathrm{G}$ and $\mathrm{I}$, we compute the attribute similarity by using the metrics that are discussed in Section~\ref{sec:metrics}. In Step~3, we label the pairs in sets $\mathrm{G}$  and $\mathrm{I}$  and add them to the training dataset. If the pair is in set $\mathrm{G}$, we label the pair as ``1'', otherwise we label it as ``0''. In Step~4, we train our model using different machine learning techniques such as linear regression and support vector machine to learn the contribution of each attribute to the profile matching attack (details of this step are discussed in Section~\ref{sec:weights}). In Step~5, the attack type is determined and profiles to be matched are selected, and hence sets $\mathrm{A_e}$ and $\mathrm{T_e}$ are constructed. For simplicity, we assume set $\mathrm{A_e}$ includes $N$ users from $A$ and set $\mathrm{T_e}$ includes $N$ users from $T$.\footnote{Sets $\mathrm{A_e}$ and $\mathrm{T_e}$ do not include any users from sets $\mathrm{A_t}$ and $\mathrm{T_t}$.} In Step~6, every profile in set $\mathrm{A_e}$  is paired with every profile in set $\mathrm{T_e}$  and the similarity between each pair is computed by using the trained model. At the last step (Step~7), profiles in sets $\mathrm{A_e}$ and $\mathrm{T_e}$ are paired by maximizing similarities using an optimization algorithm as discussed in Section~\ref{sec:hung}.

\subsection{Similarity Metrics}\label{sec:metrics}
To create the model, it is required to define similarity metrics for the attributes in $U_i^k$. In this section, we provide the details of the similarity metrics we propose for each attribute.

\subsubsection{User name similarity - $S(n^A_i,n^T_j)$}
We use Levenshtein distance~\cite{levenshtein} to calculate the similarity between user names of profiles. Given the user names $n^A_i$ and $n^T_j$ of two individuals $i$ and $j$ with profiles $U_i^A,$ and $U_j^T$, and assuming $|n^A_i|$ and $|n^T_j|$ represent the lengths of these user names, Levenshtein distance between the user names can be computed as below.
\begin{multline}
lev_{(n^A_i,n^T_j)}(|n^A_i|,|n^T_j|)= \\
\begin{cases}
max(|n^A_i|,|n^T_j|), \hspace{6.5em} \mathrm{if}\;min(|n^A_i|,|n^T_j|)= 0 \\
\\
min \begin{cases} lev_{n^A_i,n^T_j}(|n^A_i|-1,|n^T_j|)+1 \\  lev_{n^A_i,n^T_j}(|n^A_i|,|n^T_j|-1)+1 \\ lev_{n^A_i,n^T_j}(|n^A_i|-1,|n^T_j|-1)+1_{{n^A}_i\neq {n^T}_j},  &\mathrm{o/w}\end{cases}
\end{cases}\\
\label{eqn:lvt}
\end{multline}
Thus,
\begin{equation}
S(n^A_i,n^T_j)= 1- {\frac {lev_{(n^A_i,n^T_j)}(|n^A_i|,|n^T_j|)}{ max(|n^A_i|,|n^T_j|)}}.
\end{equation}

\subsubsection{Location similarity - $S(\ell^A_i,\ell^T_j)$}
Almost all OSNs have location or hometown information available in user profiles. Even if such location information is not directly available, other types of information can be used to predict location of a profile. For instance, location of a post or location-related information in the freetext can be used to predict of the location (or hometown) of a profile. Location information collected from the users' profiles is usually text-based. If the textual values of location attributes belonging to two different profiles are exactly the same, we set the location similarity of these two profiles as ``1''. Otherwise, we convert the textual information into coordinates and calculate geographic distance. We do this conversion via GoogleMaps API~\cite{googlemaps}. Let $\ell^A_i$ and $\ell^T_j$ represent the location attributes of users $i$ and $j$, respectively. We compute the location similarity between two user profiles as below.
\begin{equation}
\begin{split}
&Lat(\ell^A_i),Lon(\ell^A_i) = \mathrm{GetCoordinate}(\ell^A_i) \\
&Lat(\ell^T_j),Lon(\ell^T_j) = \mathrm{GetCoordinate}(\ell^T_j) \\
&GD=\mathrm{GeoDistance}(Lat(\ell^A_i),Lon(\ell^A_i), Lat(\ell^T_j),Lon(\ell^T_j))\\
&S(\ell^A_i,\ell^T_j) = 1- \mathrm{Norm}(GD). \nonumber
\end{split}
\label{eqn:location}
\end{equation}
Here, $\mathrm{Norm}$ is a normalization function based on all computed values in the dataset. To calculate the geographic distance ($GD$) we use haversine formula which can be computed as follows:
\begin{equation}
\begin{split}
\alpha = &sin^2(Lat(\ell^A_i)- Lat(\ell^T_j))+cos(Lat(\ell^A_i)) \cdot cos(Lat(\ell^T_j)) \\
& \cdot sin^2(Lon(\ell^A_i)-Lon(\ell^T_j))\\
\beta = &2\cdot \mathrm{atan2}(\sqrt{(\alpha)}, \sqrt{(1-\alpha)}) \\
GD &= R  \cdot \beta,
\end{split}
\label{eqn:haversine}
\end{equation}
where $R$ is earth's radius (mean radius $=6371$ km).

\subsubsection{Gender similarity - $S(g^A_i,g^T_j)$}
Availability of gender information is mostly problematic in OSNs. Some OSNs do not publicly share the gender information of their users. Furthermore, some OSNs do not even collect this information. In such cases, a profile's gender can be predicted from other information. In our model, if an OSN does not provide the gender information publicly (or does not have such information), we probabilistically infer the gender information by using a public name database. That is, we use the US social security name database\footnote{US social security name database includes year of birth, gender, and the corresponding name for babies born in the United States.} and look for a profile's name (or user name) to probabilistically infer the possible gender of the profile from the distribution of the corresponding name (among males and females) in the name database. We then use this probability as the $S(g^A_i,g^T_j)$ value between two profiles.

\subsubsection{Profile photo similarity - $S(p^A_i,p^T_j)$}
Profile photo similarity is calculated through a framework named OpenFace~\cite{amos2016openface}. OpenFace is an open source tool performing face recognition. OpenFace first detects the face (in the photo), and then preprocesses it to create a normalized and fixed-size input for the neural network. The features that characterize a person's face are extracted by the neural network and then used in classifiers or clustering techniques. OpenFace notably offers higher accuracy than previous open source frameworks. Given two profile photos $p^A_i$ and $p^T_j$, OpenFace returns the photo similarity, $S(p^A_i,p^T_j)$, as a real value between $0$ (meaning exactly the same photo) and $4$.

\subsubsection{Freetext similarity - $S(f^A_i,f^T_j)$}
Freetext data in an OSN profile could be a short biographical text or an ``about me'' page. In the proposed model, NER (named-entity recognition)~\cite{ner} is used to extract features from the freetext information. The extracted features are location, person, organization, money, percent, date, and time. To calculate the freetext similarity between the profiles of two users, we use the cosine similarity between the extracted features from each user. Let $f^A_i$ and $f^T_j$ be two vectors composed of the extracted words by the NER in the freetext provided by users $i$ and $j$ in OSNs $A$ and $T$, respectively. $\mathrm{F}_{i,j}^{A,T}$ is the set of unique words in $f^A_i$ and $f^T_j$. Let also $freq^A_i$ and $freq^T_j$ be two vectors of size $|\mathrm{F}_{i,j}^{A,T}|$ representing the frequency of each word in $\mathrm{F}_{i,j}^{A,T}$ in vectors $f^A_i$ and $f^T_j$, respectively. Then, cosine similarity between the freetexts of two user profiles can be computed as below.

\begin{equation}
\begin{split}
S(f^A_i,f^T_j) &=
\frac{\textbf{$freq^A_i$}\cdot\textbf{$freq^T_j$}}{|\textbf{$freq^A_i$}||\textbf{$freq^T_j$}|}\\
&=\frac{{\sum_{i=1}^{|\mathrm{F}_{i,j}^{A,T}|}{freq^A_i}{freq^T_j}}}{\sqrt{\sum_{i=1}^{|\mathrm{F}_{i,j}^{A,T}|}{(freq^A_i)}^{2}\sum_{j=1}^{|\mathrm{F}_{i,j}^{A,T}|}{(freq^T_j)}^{2}}}\label{eqn:CosineSim}.
\end{split}
\end{equation}

\subsubsection{Activity pattern similarity - $S(a^A_i,a^T_j)$}
Activity pattern similarity is defined as the similarity between observed activity patterns of two profiles (e.g., login or post). Let $a^A_i$ represent a vector including the times of last $|a^A_i|$ activities of user $i$ in OSN $A$. Similarly, $a^T_j$ is a vector including the times of last $|a^T_j|$ activities of user $j$ in OSN $T$. First, we compute the time difference between every entry in $a^A_i$ and $a^T_j$ and we determine $\min(|a^A_i|,|a^T_j|)$ pairs whose time difference is the smallest. Then, we compute the normalized distance between these $\min(|a^A_i|,|a^T_j|)$ pairs to compute the activity pattern similarity between two profiles.

\subsubsection{Interest similarity - $S(t^A_i,t^T_j)$}
OSNs provide a platform in which users share their opinions via posts (e.g., tweets or tips), and this shared content is composed of different topics. In our framework, we use the topic similarity of shared content (posts) in profiles to calculate the interest similarity between the users. In highlevel, first, we create a topic model using the posts of randomly selected users from both the auxiliary and the target OSNs. Then, by using the created model, we compute the topic distribution of each post generated by the users of the auxiliary and the target OSNs. Finally, we compute the interest similarity from the distance of the computed topic distributions.

To create a topic model we use Latent Dirichlet Allocation (LDA)~\cite{Blei:2003:LDA:944919.944937}. LDA  is a generative probabilistic and graphical model. In LDA it is assumed that each word is drawn from one of the hidden topics and each document (shared content) is a mixture of multiple topics where the topics are modeled as probability distributions over a fixed vocabulary. Thus, LDA aims to infer the document-topic, topic-word distributions and topic assignments to words.

Let $t^A_i$ and $t^T_j$ represent the set of posts generated in two profiles $U^A_i$ and $U^T_j$. Assume a topic model consisting $\theta$ topics ($LDA$) is created. Then, we first find the normalized topic distribution of each profile as follows:
\begin{equation*}
\begin{split}
d_{t^A_i} &=\frac{1}{|t^A_i|}\sum_{m=1}^{|t^A_i|}LDA(t^A_i(m))\\
d_{t^T_i} &=\frac{1}{|t^T_i|}\sum_{m=1}^{|t^T_i|}LDA(t^T_i(m)),\\
\end{split}
\end{equation*}
where both $d_{t^A_i}$ and $d_{t^T_j}$ are vectors of size $\theta$ representing the distributions of the topics the users posted about. Then, we find the interest similarity between users $i$ and $j$ as follows:
\begin{equation}
S(t^A_i,t^T_j)=1-\frac{1}{\theta}\sum_{m=1}^{\theta}|d_{t^A_i}(m)-d_{t^T_j}(m)|.
\end{equation}

\subsubsection{Sentiment similarity - $S(s^A_i,s^T_j)$}
Users typically express their emotions when sharing their opinions about certain issues on OSNs. To determine whether the shared text (e.g., post, tip, or tweet) expresses positive or negative sentiment we use sentiment analysis through Python NLTK (natural language toolkit) Text Classification~\cite{toolkit}. Given the text to analyze, the sentiment analysis tool returns the probability for positive and negative sentiment in the text.  Since users' mood is affected from different factors, it is realistic to assume that it might change by time (e.g., daily). Thus, we compute the daily sentiment profile of each user, and daily sentiment similarity between the users. For this, first, we compute the normalized distribution of the positive and negative sentiments per day for each user, and then we find the normalized distance between these distributions for each user pair.

\subsection{Training the Model}\label{sec:weights}

As discussed, we first construct sets $\mathrm{A_t}$ and $\mathrm{T_t}$ for training. Also, set $\mathrm{G}$ includes pairs of profiles $(U_i^A,U_j^T)$ that belong to the same individual and set $\mathrm{I}$ includes pairs of profiles $(U_i^A,U_j^T)$ from $\mathrm{A_t}$ and $\mathrm{T_t}$ that belong to different individuals. We refer to the pairs in $\mathrm{G}$ as ``coupled profiles'' and the ones in $\mathrm{I}$ as ``uncoupled profiles''. We first compute the individual attribute similarities between each pair of coupled and uncoupled  profiles in $\mathrm{G}$ and $\mathrm{I}$ using the similarity metrics described in Section~\ref{sec:metrics}. Then, to train our model, we use two different machine learning techniques: (i) linear regression and (ii) support vector machine (SVM).

\subsubsection{Linear Regression}\label{sec:linearregression}

Prior work shows that linear regression is an efficient way to learn attribute weights for social network related data~\cite{debnath2008feature}. Therefore, we apply linear regression to our training dataset in order to learn the contribution (or weight) of each attribute for the profile matching attack. Notations we use for this part are listed in Table~\ref{table:model}.
\begin{table}[h]
	\centering
	\resizebox{0.20\textwidth}{!}{
		\begin{tabular}{|c|c|}
			\hline
			\(w_n\) & Weight of user name\\
			\hline
			\(w_\ell\)  & Weight of location\\
			\hline
			\(w_g\) & Weight of gender\\
			\hline
			\(w_p\) & Weight of profile photo\\
			\hline
			\(w_f\) & Weight of freetext\\
			\hline
			\(w_a\) & Weight of activity pattern\\
			\hline
			\(w_t\) & Weight of interest\\
			\hline
			\(w_s\) & Weight of sentiment\\
			\hline
		\end{tabular}
	}
	\caption{Notations used in the linear regression.}
	\label{table:model}
	\vspace{-15pt}
\end{table}

We learn the contribution (or weight) of each attribute for the profile matching attack via linear regression as follows:
\begin{equation}
S(U_i^A,U_j^T )  =
\begin{cases}
1=y, & \text{if $(U_i^A,U_j^T)\in\mathrm{G}$}  \\
0=y, & \text{if $(U_i^A,U_j^T)\in\mathrm{I}$}
\end{cases}
\label{eqn:Linearregsss}
\end{equation}
where,
\begin{equation}
\begin{split}
\centering
y =& w_0 + w_n \times S(n_i^A,n_j^T) + w_\ell \times  S(\ell_i^A,\ell_j^A) \\
&+ w_g \times  S(g_i^A,g_j^T) + w_p \times  S(p^A_i,p^T_j )  \\
&+ w_f \times  S(f_i^A,f_j^T) + w_a \times  S(a^A_i,a^T_j ) + w_t \times S(t_i^A,t_j^T)
\end{split}
\label{eqn:similarity}
\end{equation}

\subsubsection{Support Vector Machine}\label{sec:SVM}

One of the recent works~\cite{liu2014hydra, liuarticle} presents a profile matching framework that relies on support vector machine(SVM).
Thus, to observe the effect of different machine learning techniques to the success of the attack, we also applied SVM for training part of proposed attack. SVMs use decision planes and they are mainly used for classification tasks. However, they can be also used for regression tasks (as in our work). In SVM regression, the input is mapped onto an $m$-dimensional feature space ($m$ is the number of attributes in our case), and then a linear model is constructed as follows:
\begin{equation}
\centering
S(U_i^A,U_j^T ) = f(x,w)= \sum_{j=1}^{m} w_j g_j(x)+b,
\end{equation}
where $x$ is an $m$-sized vector that is composed individual attribute similarities for each profile pair, $g_j(x)$ ($j=1,\ldots,m$) is a set of nonlinear transformations, and $b$ is the bias term. Thus, to learn the weights ($w$) and the bias parameter ($b$) we minimize the following:
\begin{equation}
\begin{split}
\centering
\frac{1}{2}\|w\|^2 + C\sum_{i=1}^n \xi_{ii^`}\nonumber
\end{split}
\label{eqn:objective}
\end{equation}
\begin{equation}
\centering
s.t.
\begin{cases}
& \text{$y_i -f(x_i,w) \le \varepsilon + \xi^*_i $}  \\
& \text{$f(x_i,w)-y_i \le \varepsilon+ \xi_i $}  \\
& \text{$\xi_i \xi^*_i \geq 0 $},
\end{cases}
\end{equation}
where $y_i$ equals to $1$ if the pair is coupled, otherwise $0$. $\varepsilon$ is the empirical risk of regression, $\xi$ is slack variable, and $n$ is the number of slack variables.

\subsection{Matching Profiles}\label{sec:hung}
As discussed, for profile matching attack, we consider the users in sets $\mathrm{A_e}$ and $\mathrm{T_e}$ from the auxiliary and the target OSNs. For simplicity, we also assume that both sets include $N$ users.\footnote{The case when the sizes of the OSNs are different can be also handled similarly (by padding one OSN with dummy users to equalize the sizes).} Before the actual profile matching, individual attribute similarities between every profile in $\mathrm{A_e}$ and in $\mathrm{T_e}$ are computed using the similarity metrics described in Section~\ref{sec:metrics}. Then, the general similarity $S(U^{A}_i, U^{T}_j)$  is computed between every user in $\mathrm{A_e}$ and $\mathrm{T_e}$ using the weights determined in Section~\ref{sec:weights}. Let $Z$ be a $N \times N$ similarity matrix that is constructed from the pairwise similarities between the users in $\mathrm{A_e}$ and $\mathrm{T_e}$ as below:
\begin{equation*}
Z =
\begin {bmatrix}
S(U^A_0,U^T_0)&  \ldots& S(U^A_0,U^T_N)\\
\vdots& \ddots& \vdots\\
S(U^A_N,U^T_0)& \ldots& S(U^A_N,U^T_N)
\end{bmatrix}
\end{equation*}
Our goal is to obtain a one-to-one matching between the users in $\mathrm{A_e}$ and $\mathrm{T_e}$ that would also maximize the total similarity. To achieve this matching, we use the Hungarian algorithm, a combinatorial optimization algorithm that solves the assignment problem in polynomial time~\cite{hungarian}. The objective function of the Hungarian algorithm can be expressed as below.
\begin{equation*}
max \sum\limits_{i=1}^N \sum\limits_{j=1}^N {{-Z}_{ij}x_{ij}},
\end{equation*}
where, $-Z_{ij}$ represents the similarity between $U^A_i$ and $U^T_j$ (i.e., $S(U^A_i,U^T_j)$). Also, $x_{ij}$ is a binary value, that is, $x_{ij}=1$ if profiles $U^A_i$ and $U^T_j$  are matched as a result of the algorithm, and $x_{ij}=0$ otherwise. After performing the Hungarian algorithm to the $Z$ matrix, we obtain a matching between the users in $\mathrm{A_e}$ and $\mathrm{T_e}$ that maximizes the total similarity. Note that we multiply $Z_{ij}$ values with -1, in order to obtain the maximum similarity (profit).

\section{Evaluation}\label{sec:evaluation}
In this section, we evaluate the proposed framework by using real data from three OSNs. We also study the impact of various sets of attributes to the profile matching.

\subsection{Data Collection}\label{sec:datasetcreate}

In the literature there are limited datasets that can be used for profile matching between unstructured OSNs. Thus, to evaluate our proposed framework, we created two dataset that consist of users from three OSNs (Twitter, Foursquare, and Google+) with several attributes. The most challenging part of data collection was to obtain the ``coupled'' profiles between OSNs that belongs to same person in real-life.
In the following, we discuss our data collection methodology.

\subsubsection{Dataset~1: Twitter - Foursquare}

We utilized Twitter API (through twitter4j Java library~\cite{twitter4j}), and Foursquare API~\cite{foursquareapi} for the data collection (i.e., collection of the attributes for the coupled and uncoupled profiles). To automate the coupled profile collection process, we used Twitter Streaming API~\cite{twitterstreaming}. When an individual generates a check-in in the Swarm app\footnote{Swarm is a companion app to Foursquare. It allows users to share their locations within their social network.}~\cite{swarm}, and publishes it via Twitter with a tweet, we catch the corresponding tweet, parse Twitter and Foursquare attributes, and connect the corresponding Twitter and Foursquare accounts to each other (as coupled profiles). We used Foursquare as our auxiliary OSN ($A$) and Twitter as our target OSN ($T$).

Furthermore, we also randomly paired uncoupled profiles which are used for training and testing the proposed algorithm. From each coupled or uncoupled profile, we extracted the following information: (i) from Foursquare; firstname/lastname, last 20 tips of the user (freetext comments about the locations), times of last 20 tips, freetext about the user, gender, location, and profile photo, and (ii) from Twitter; name/screenname, last 20 tweets of the user and their times, freetext about the user, location, profile photo, and timezone.

Due to the special characteristics of Twitter and Foursquare, we slightly changed some of our similarity calculations as follows. On Twitter, users can declare a ``screen name'' and ``name'' separately. We observed that some users declare their real first names or last names in either of these fields. On the other hand, on Foursquare, users fill the ``firstname'' and ``lastname'' fields. As a result, we created 4 types of name similarity metrics by using combination of names: (i) Twitter screen name and Foursquare firstname, (ii) Twitter screen name and Foursquare lastname, (iii) Twitter name and Foursquare firstname, (iv) Twitter name and Foursquare lastname.

\subsubsection{Dataset~2: Google+ - Twitter Dataset}

In this dataset, we used Twitter as our auxiliary OSN ($A$) and Google+ as our target OSN ($T$). To create this dataset, we used Google+ and Twitter APIs~\cite{googleapi,twitter4j}. To collect the coupled profiles between Google+ and Twitter, we first visited random Google+ profiles and parsed the ``social links data'' (that includes the URLs to the other OSN accounts of the users) of each visited profile. If a social link data contains a valid Twitter profile, we then visited the corresponding Twitter profile and extracted the same attributes as in Database~1 from the Twitter profile. Also, we extracted the following information from the Google+ profiles of the users: given name, last name, username, location, profile photo, gender, skills, tagline, about me, the last 20 activities of the user, and the times of the last 20 activities. 

Similar to Database~1, due to special characteristics of Google+ and Twitter, we created 6 types of name similarity metrics. As mentioned, in Twitter users can share their ``screen name'' and ``name''. On the other hand, in Google+, users may share their ``given name'', ``last name'', and ``username''. Thus, we obtained 6 different combinations for the user name attribute: (i) Twitter screen name and Google+ given name, (ii) Twitter screen name and Google+ last name, (iii) Twitter screen name and Google+ username, (iv) Twitter name and Google+ given name, (v) Twitter name and Google+ last name, and (vi) Twitter name and Google+ username.

\begin{figure*}
	\centering
	\subfigure[User Name]{
		\includegraphics[scale=0.3]{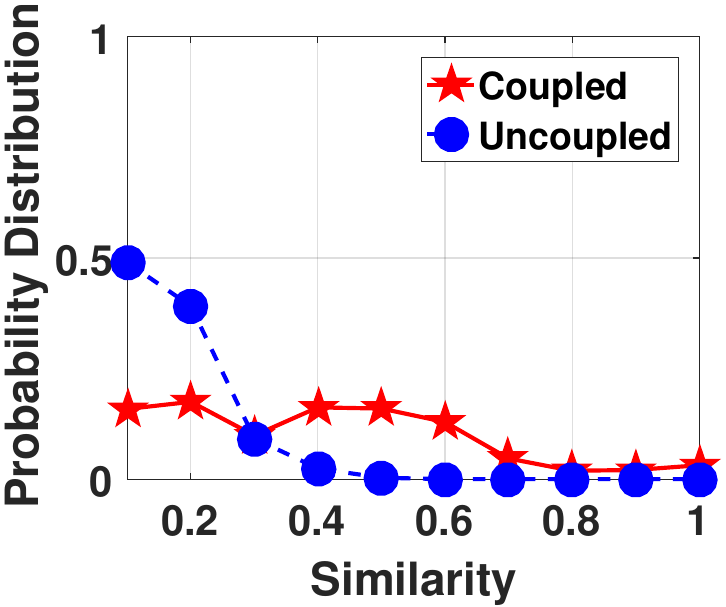}
	}
	\subfigure[Location]{
		\includegraphics[scale=0.3]{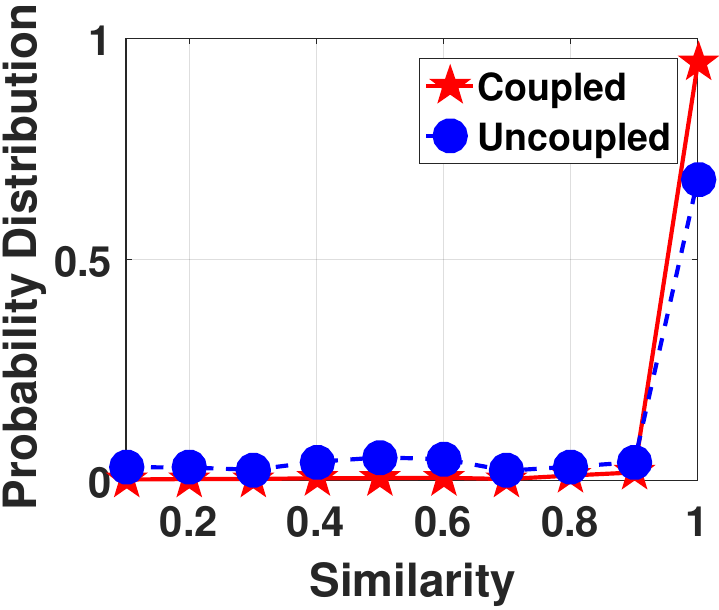}
	}
	\subfigure[Gender]{
		\includegraphics[scale=0.3]{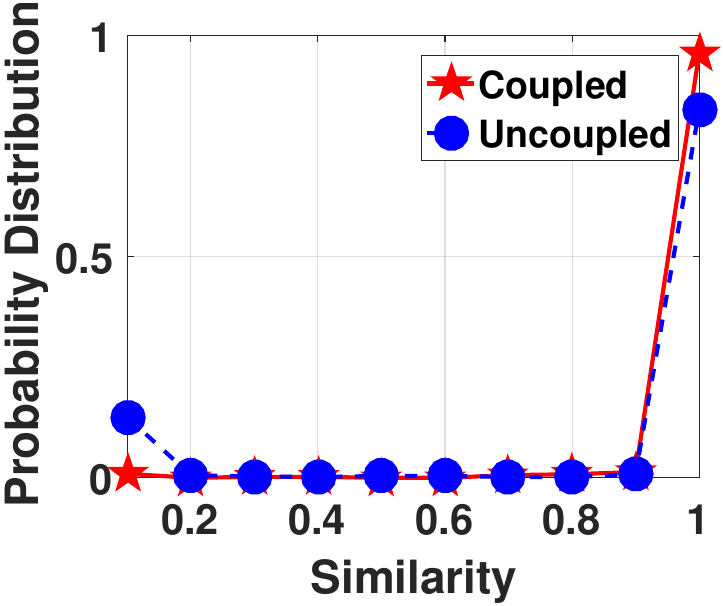}
	}
	\subfigure[Profile Photo]{
		\includegraphics[scale=0.3]{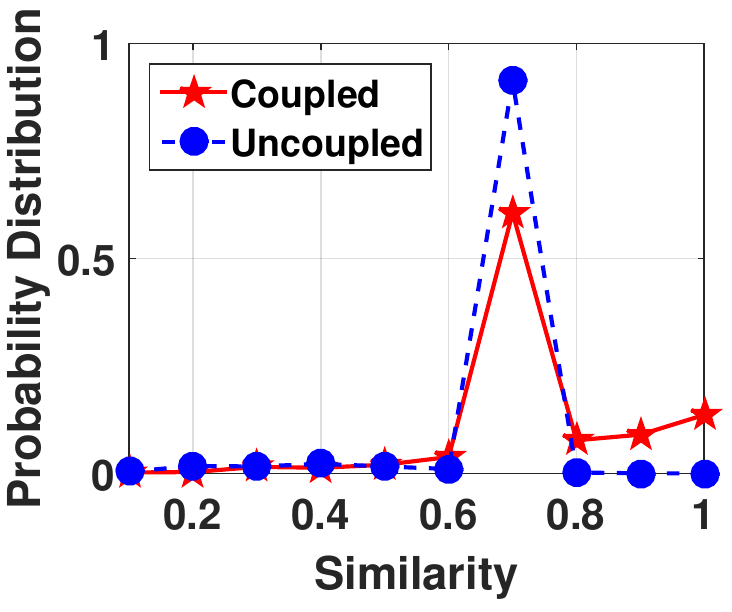}
	}
	\subfigure[Freetext]{
		\includegraphics[scale=0.3]{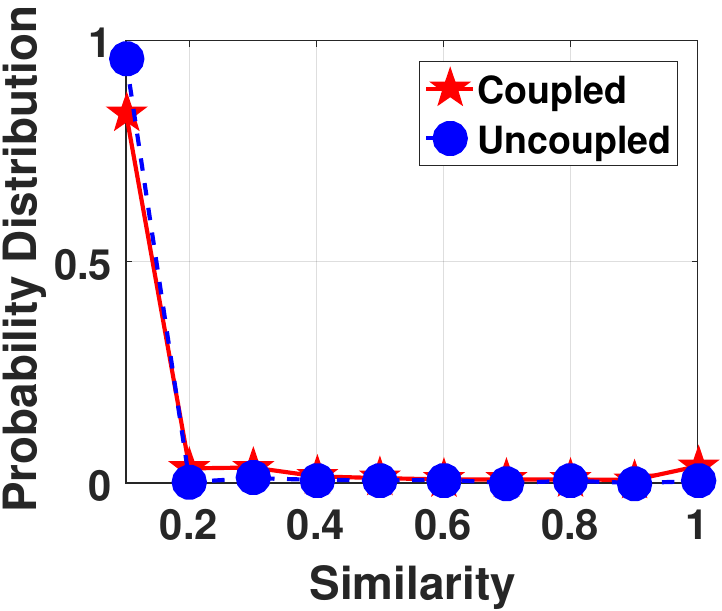}
	}
	\subfigure[Activity Pattern]{
		\includegraphics[scale=0.3]{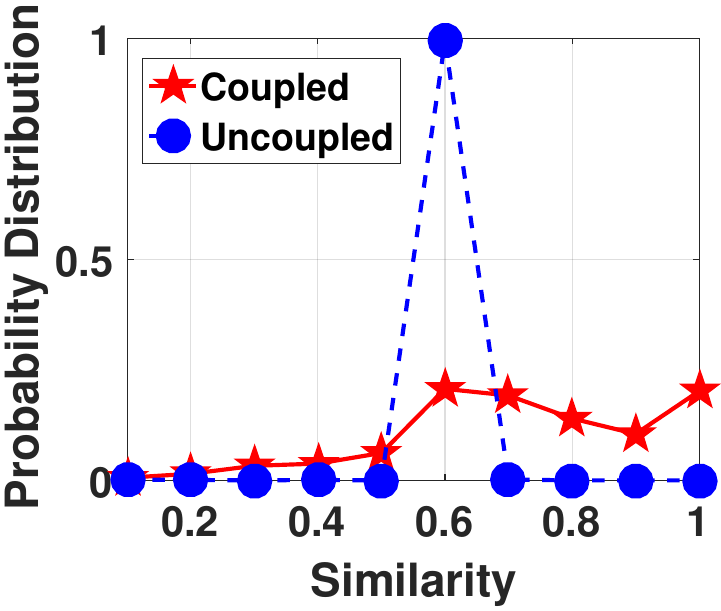}
	}
	\subfigure[Interest]{
		\includegraphics[scale=0.3]{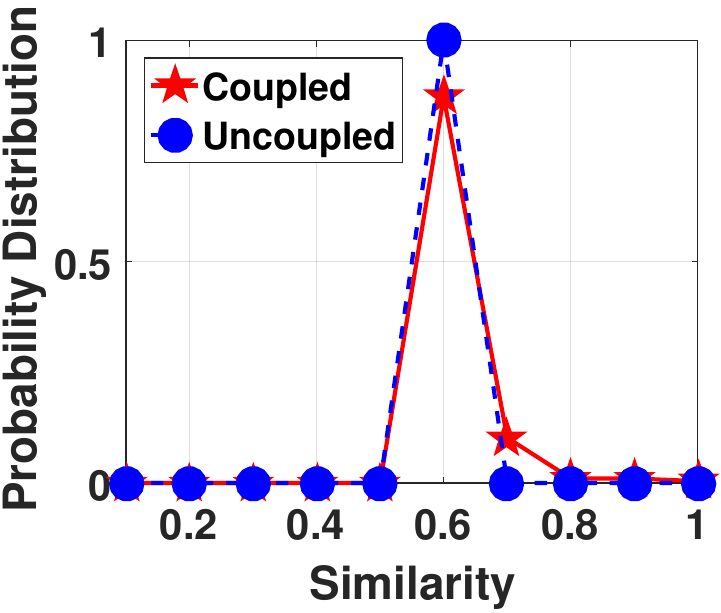}
	}
	\subfigure[Sentiment]{
		\includegraphics[scale=0.3]{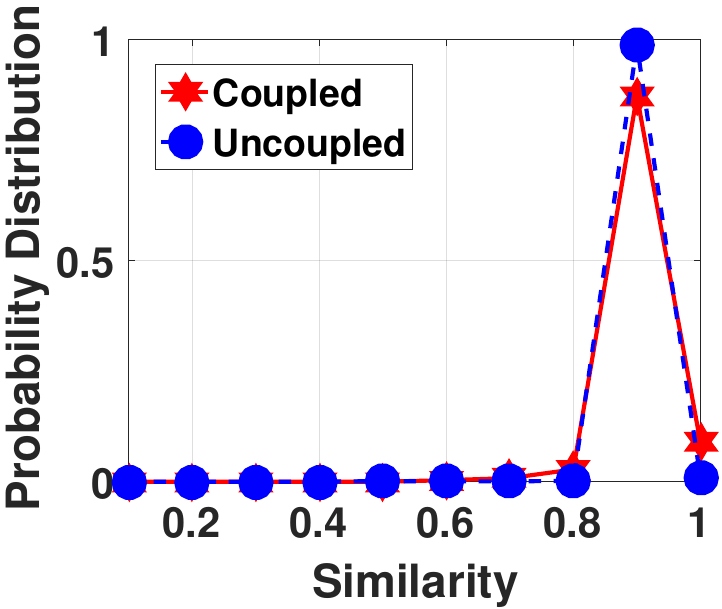}
	}
	\caption{Probability distributions of attribute similarities among the pairs used for the training in the Twitter - Foursquare dataset ($\mathrm{D1}$).}
	\label{fig:dist}
\end{figure*}

\begin{figure*}
	\centering
	\subfigure[User Name]{
		\includegraphics[scale=0.3]{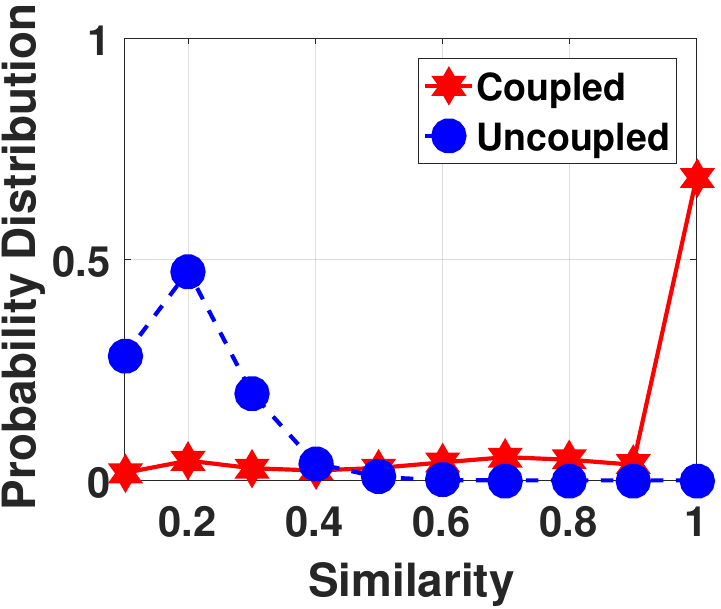}
	}
	\subfigure[Location]{
		\includegraphics[scale=0.3]{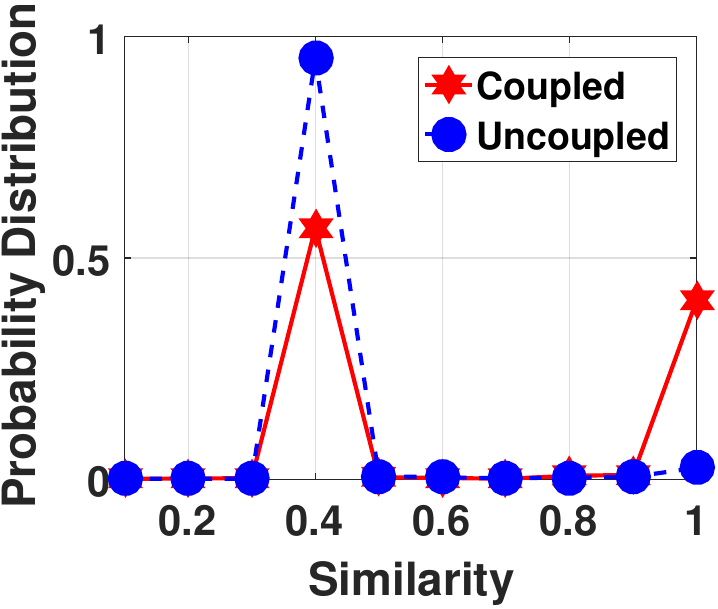}
	}
	\subfigure[Gender]{
		\includegraphics[scale=0.3]{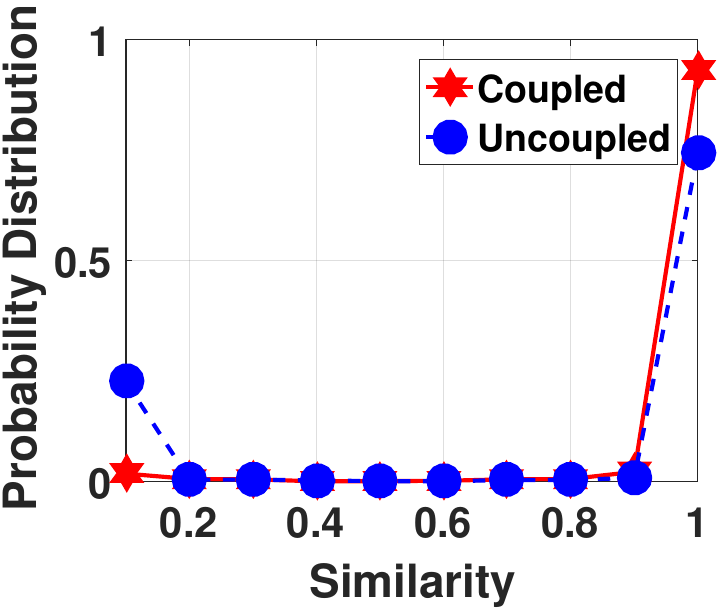}
	}
	\subfigure[Profile Photo]{
		\includegraphics[scale=0.3]{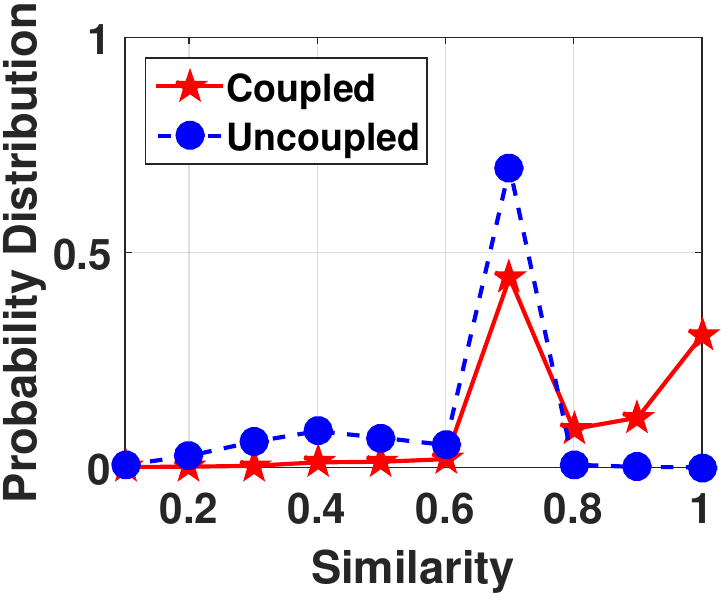}
	}
	\subfigure[Freetext]{
		\includegraphics[scale=0.3]{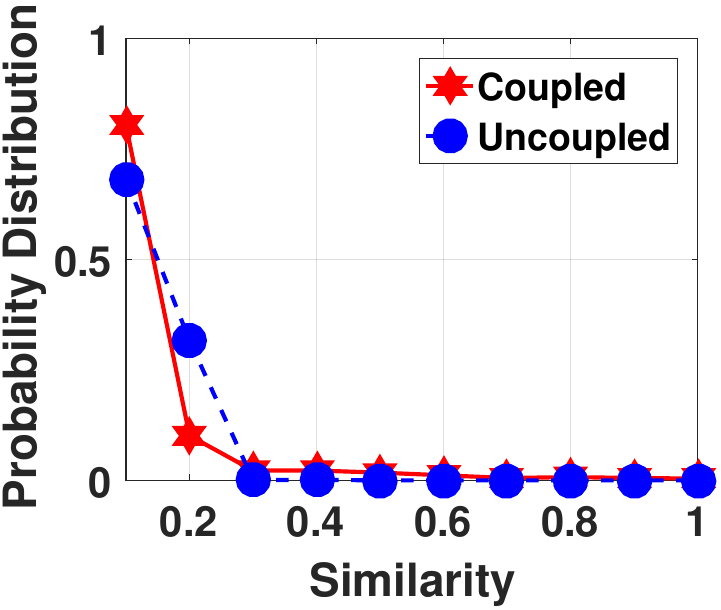}
	}
	\subfigure[Activity Pattern]{
		\includegraphics[scale=0.3]{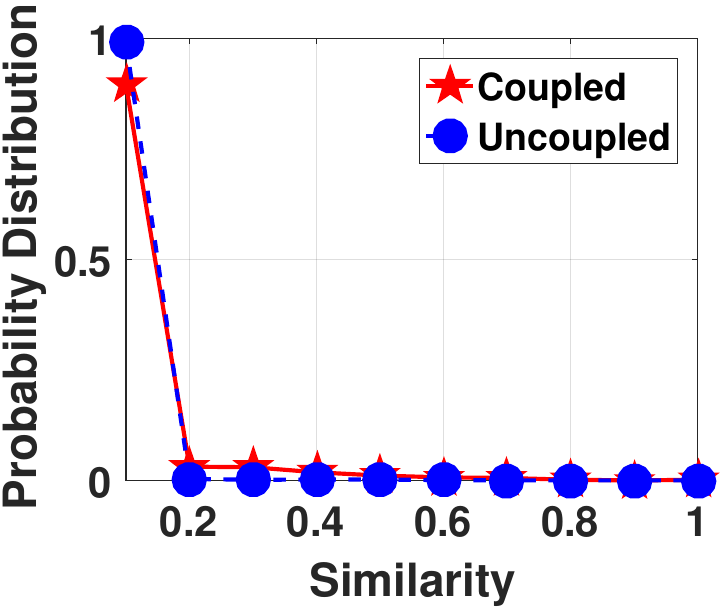}
	}
	\subfigure[Interest]{
		\includegraphics[scale=0.3]{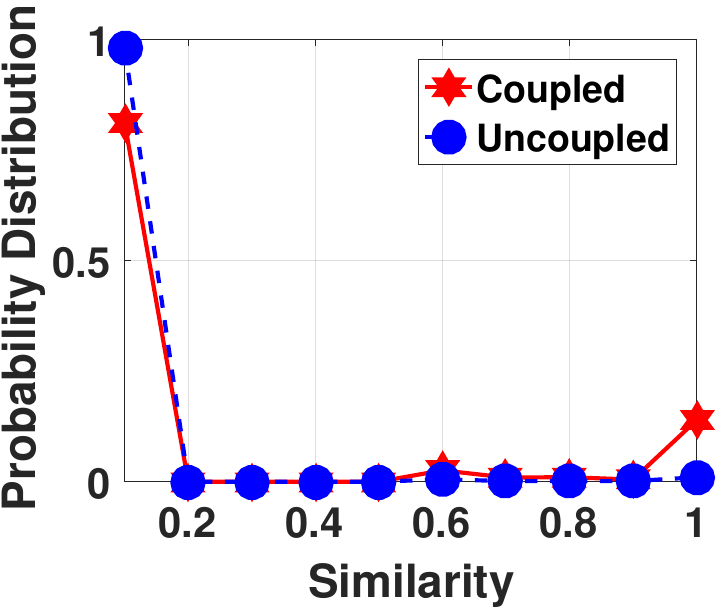}
	}
	\subfigure[Sentiment]{
		\includegraphics[scale=0.3]{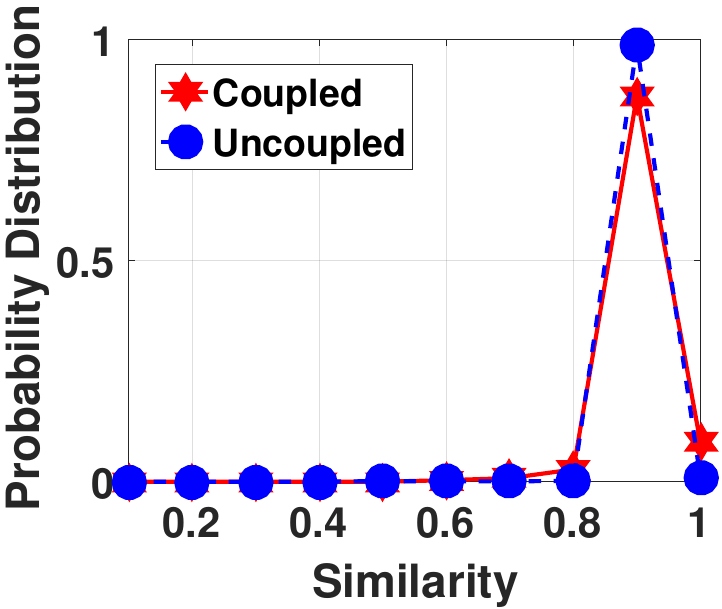}
	}
	\caption{Probability distributions of attribute similarities among the pairs used for the training in the Google+ - Twitter dataset ($\mathrm{D2}$).}
	\label{fig:dist_1}
\end{figure*}

As a result, both Dataset~1 and Dataset~2 consists of 4000 profile pairs each, of which 2000 are coupled and 2000 are uncoupled profiles. To create the LDA model, we randomly sampled a total of 15000 tweets (from Twitter), tips (from Foursquare), and posts (from Google+) and generated the model by using this data. Then, we applied the model to the posts of the users to find the interest similarity as discussed in Section~\ref{sec:metrics}. Note that there may be missing attributes (that are not published by the users) in the database. Due to such missing attributes, it is not possible to compute the similarity values for some attributes between some users. In such cases, based on the distributions of the similarity values of each attribute between the coupled and uncoupled pairs (as also shown in Figures~\ref{fig:dist} and~\ref{fig:dist_1}), we assign a value for the similarity that minimizes both the false positive and false negative probabilities. 

\subsection{Training}\label{sec:learning}

In the rest of the paper, we will hold the discussion over a target and auxiliary network as the training process is the same for both datasets. As mentioned, in our first dataset, Twitter is the auxiliary network and Foursquare is the target network. In the second dataset, Google+ is the target network and Twitter is the auxiliary network. From each dataset, we selected 3000 profile pairs for training. These pairs consists of 1500 coupled and 1500 uncoupled profile pairs. Probability distributions of the attribute similarities for the training instances in both datasets are shown in Figures~\ref{fig:dist} and~\ref{fig:dist_1}.

To train the model, we used two different machine learning techniques: (i) linear regression and (ii) support vector machine. Overall, we conducted four experiments by using different sets of attributes we considered. For each experiment, we also present the weights obtained as a result of the linear regression model in Table~\ref{table:weights1} as those provide a good representation for the importance of the attributes for the profile matching attack. As discussed in Section~\ref{sec:datasetcreate}, we considered multiple user name similarity metrics for both datasets. In Table~\ref{table:weights1}, we only report the user name similarity for the metric with the highest weight (for the linear regression model) for each dataset.

In our first experiment ($\mathrm{Experiment~1}$), we used all the attributes (including all user name similarity metrics) we extracted from both OSNs for the training. We observe that location, user name (in particular Foursquare firstname - Twitter screen name in Dataset~1 and Twitter name - Google+ username in Dataset~2), and profile photo attributes are the most identifying attributes to determine whether two profiles belong to same individual or not. In the second experiment ($\mathrm{Experiment~2}$), we only used the four most identifying attributes based on the weights in Table~\ref{table:weights1} for $\mathrm{Experiment~1}$. Thus, we only used the attributes related to the user name (i.e., the user name similarity for the metric with the highest weight), location, profile photo, gender (for Dataset~2), and activity pattern (for Dataset~1).

We observed that the most identifying attribute for the profile matching attack is the user name. Since it is trivial for the users to obtain a different user name for their accounts in different OSNs, for the remaining experiments, we did not use the user name attribute and we evaluated the success of other (less identifying) attributes for the attack. Thus, in the third experiment ($\mathrm{Experiment~3}$), we used all the attributes but the user name. In the fourth and the last experiment ($\mathrm{Experiment~4}$), we only considered the activity patterns, freetext, interests (that is extracted from users' posts), and sentiment. Note that this last scenario can be also used to quantify the risk of profile matching between an OSN and a profile in a forum (in which users typically remain anonymous, and activity patterns, freetext, interests, and sentiment are the only attributes that can be learnt about the users). Following our training model, we compute the general similarity between profiles $U_i^A$ and $U_j^T$ for both machine learning techniques.
\begin{table}[h]
	\centering
	\resizebox{0.5\textwidth}{!}{
		\begin{tabular}{c|c|c|c|c|c|c|c|c|}
			\cline{2-9}
			& \multicolumn{2}{|c|}{$\mathrm{Experiment~1}$} &  \multicolumn{2}{|c|}{$\mathrm{Experiment~2}$} & \multicolumn{2}{|c|}{$\mathrm{Experiment~3}$} & \multicolumn{2}{|c|}{$\mathrm{Experiment~4}$}\\
			\hline
			\multicolumn{1}{|c|}{Attribute} & W($\mathrm{D1}$) & W($\mathrm{D2}$) & W($\mathrm{D1}$) & W($\mathrm{D2}$) & W($\mathrm{D1}$) & W($\mathrm{D2}$) & W($\mathrm{D1}$) & W($\mathrm{D2}$) \\
			\hline
			\multicolumn{1}{|c|}{user name ($w_n$)}  & $0.65$ & $0.56$ & $0.93$ & $0.75$ & N/A & N/A & N/A & N/A\\
			\hline
			\multicolumn{1}{|c|}{location ($w_{\ell}$)}  & $0.35$ & $0.13$ & $0.45$ & $0.17$ & $0.56$ & $0.55$ & N/A & N/A\\
			\hline
			\multicolumn{1}{|c|}{gender ($w_g$)} & $0.21$ & $0.09$ & N/A & $0.10$ & $0.34$ & $0.30$ & N/A & N/A\\
			\hline
			\multicolumn{1}{|c|}{profile photo ($w_p$)}  & $0.35$  & $0.26$ & $0.51$  & $0.28$ & $0.83$ & $1.06$ & N/A & N/A\\
			\hline
			\multicolumn{1}{|c|}{freetext ($w_f$)}  & $0.06$ & $0.06$ & N/A & N/A & $0.14$ & $0.24$ & $0.22$ & $0.45$\\
			\hline
			\multicolumn{1}{|c|}{activity pattern ($w_a$)}  & $0.63$ & $\sim0$ & $0.88$ & N/A & $1.10$ & $0.22$ & $1.3$ & $0.26$\\
			\hline
			\multicolumn{1}{|c|}{interest ($w_t$)}  & $\sim0$ & $0.06$ & N/A & N/A & $\sim0$ & $0.24$ & $\sim0$ & $0.45$\\
			\hline
			\multicolumn{1}{|c|}{sentiment ($w_s$)}  & $\sim0$ & $\sim0$ & N/A & N/A & $\sim0$ & $0.15$ & $\sim0$ & $\sim0$\\
			\hline
		\end{tabular}
	}
	\caption{Weights of attributes computed as a result of linear regression model for $\mathrm{Experiment~1}$ (in which we use all the attributes),  $\mathrm{Experiment~2}$ (in which we only use the four most identifying attributes), $\mathrm{Experiment~3}$ (in which we use all the attributes except for the user name), and $\mathrm{Experiment~4}$ (in which we use the weakest identifiers) for both Dataset~1 ($\mathrm{D1}$) and Dataset~2 ($\mathrm{D2}$). N/A means the corresponding attribute is not used in the corresponding experiment.}
	\label{table:weights1}
	\vspace{-15pt}
\end{table}

\subsection{Profile Matching and Evaluation Metrics}\label{sec:baseline}

After training the model for all the experiments, we selected 1000 users from the auxiliary OSN and 1000 users from the target OSN to construct sets $\mathrm{A_e}$ and $\mathrm{T_e}$, respectively (for both datasets). Note that none of these users were involved in the training set. Among these profiles we had 500 coupled pairs. Thus, we evaluated the success of our proposed framework based on the matching between these coupled profiles.

To evaluate our model, we consider two types of profile matching attacks: (i) targeted attack, and (ii) global attack. In targeted attack, the goal of the attacker is to match the anonymous profiles of one or more target individuals from $T$ to their corresponding profiles in $A$. In the global attack, the goal of the attacker is to match all profiles in $\mathrm{A_e}$ to all profiles in $\mathrm{T_e}$. In other words, the goal is to deanonymize all anonymous users in the target OSN (who has accounts in the auxiliary OSN).

In both targeted and global attacks, we use the Hungarian algorithm for profile matching between the auxiliary and the target OSN (as discussed in Section~\ref{sec:hung}). Hungarian algorithm provides a one-to-one match between all the users in the auxiliary and the target OSN. However, we cannot expect that all anonymous users in the target OSN to have profiles in the auxiliary OSN (we are only interested in the ones that have profiles in both OSNs). Therefore, some matches provided by the Hungarian algorithm are useless for us. Thus, we define a confidence value and we only consider the matches that are above the confidence value to evaluate the precision and recall of our framework. For this purpose,  we set a ``similarity threshold''. We examine the general similarity values between the matches as a result of the Hungarian algorithm and we only consider the matches that are above the similarity threshold to compute the precision and recall.

In the following, we summarize the notations we use to compute precision and recall for the targeted attack.
\begin{itemize}
	\item True positive ($tp$): Both profiles from a coupled profile are correctly paired with each other above the similarity threshold.
	\item False positive ($fp$): One of the profiles from a coupled profile pair is incorrectly paired with a different profile above the similarity threshold.
	\item True negative ($tn$): One of the profiles from a coupled profile pair is paired with a different profile below the similarity threshold.
	\item False negative ($fn$): Coupled profiles are paired below the similarity threshold.
\end{itemize}
For the global attack, in addition to the above rules, (i) false positive also includes uncoupled profile pairs that are paired above the similarity threshold, and (ii) true negative also includes the uncoupled profile pairs that are paired below the similarity threshold. Then, we compute the precision and recall as $tp/(tp+fp)$ and $tp/(tp+fn)$, respectively. We also evaluate the ``success rate'' of the attack as the fraction of correctly matched coupled pairs to all coupled pairs regardless of the similarity threshold.

In all experiments, we compare our proposed attack with the baseline approach for profile matching as described in the following. In the baseline approach, we only use the obvious identifiers such as user name, location, gender, and profile photo (we use our proposed metrics to compute the individual similarities of these attributes). K-nearest neighbor (KNN), decision tree, and SVM techniques are used to classify the pairs as coupled or uncoupled. In KNN, a pair is assigned to the most common class among its k-nearest neighbors. A decision tree has a tree like structure in which each internal node represents a ``test'' on a feature, each branch represents the result of the test, and each leaf represents a class label. In SVM model the training data are represented as points in space and the data of different categroies are divided by a clear gap. New examples are mapped into the same space and are classified by checking on which side of the gap they fall. To implement this baseline approach, first, we train the classifiers with the training dataset constructed in Section~\ref{sec:weights} (including only user name, location, gender, and profile photo). Then, based on the trained model, each new pair is classified using either KNN, decision tree, or SVM algorithm.

\subsection{Results}\label{sec:results}

\begin{figure*}
	\centering
	\begin{minipage}{0.45\textwidth}
		\centering
		\begin{subfigure}[Global Attack]{\includegraphics[scale=0.19]{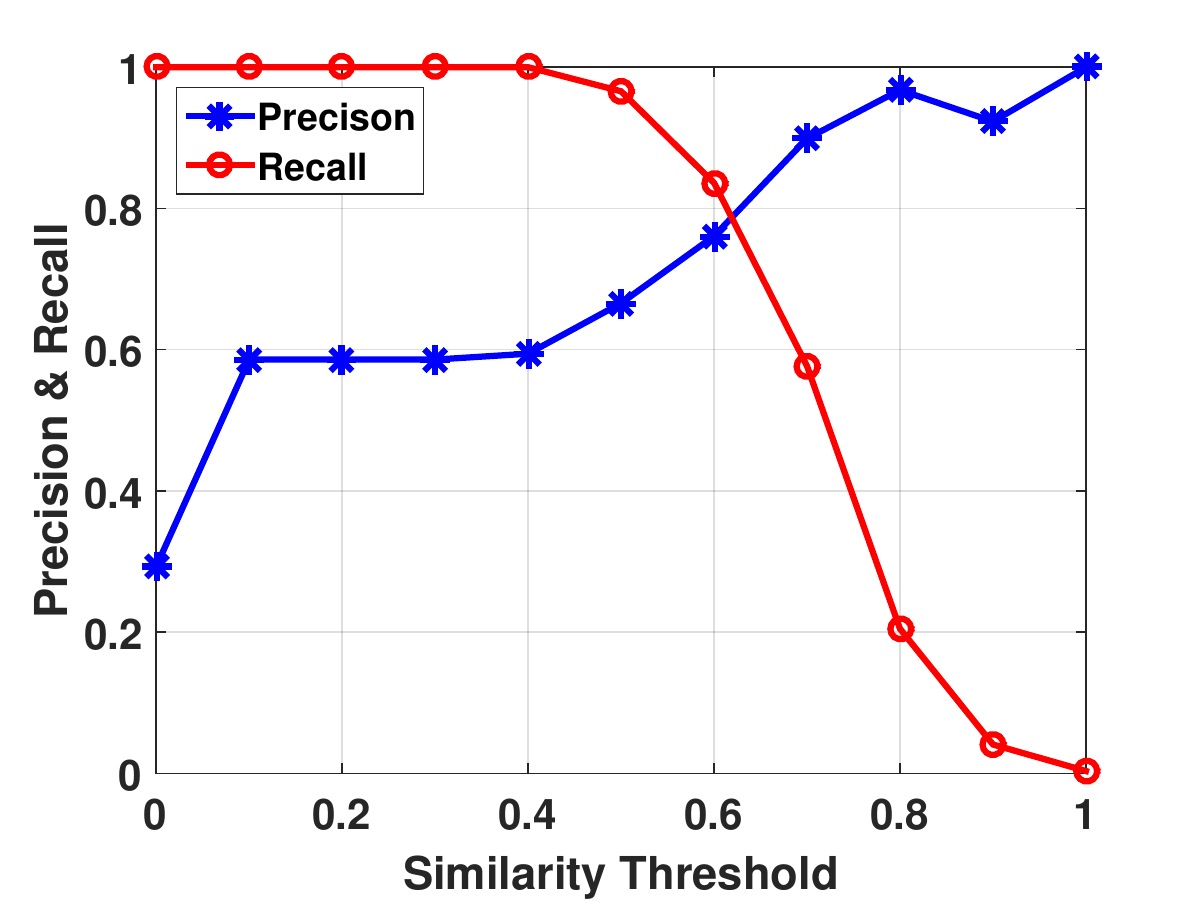}}
		\end{subfigure}\hfill
		\begin{subfigure}[Targeted Attack]{\includegraphics[scale=0.19]{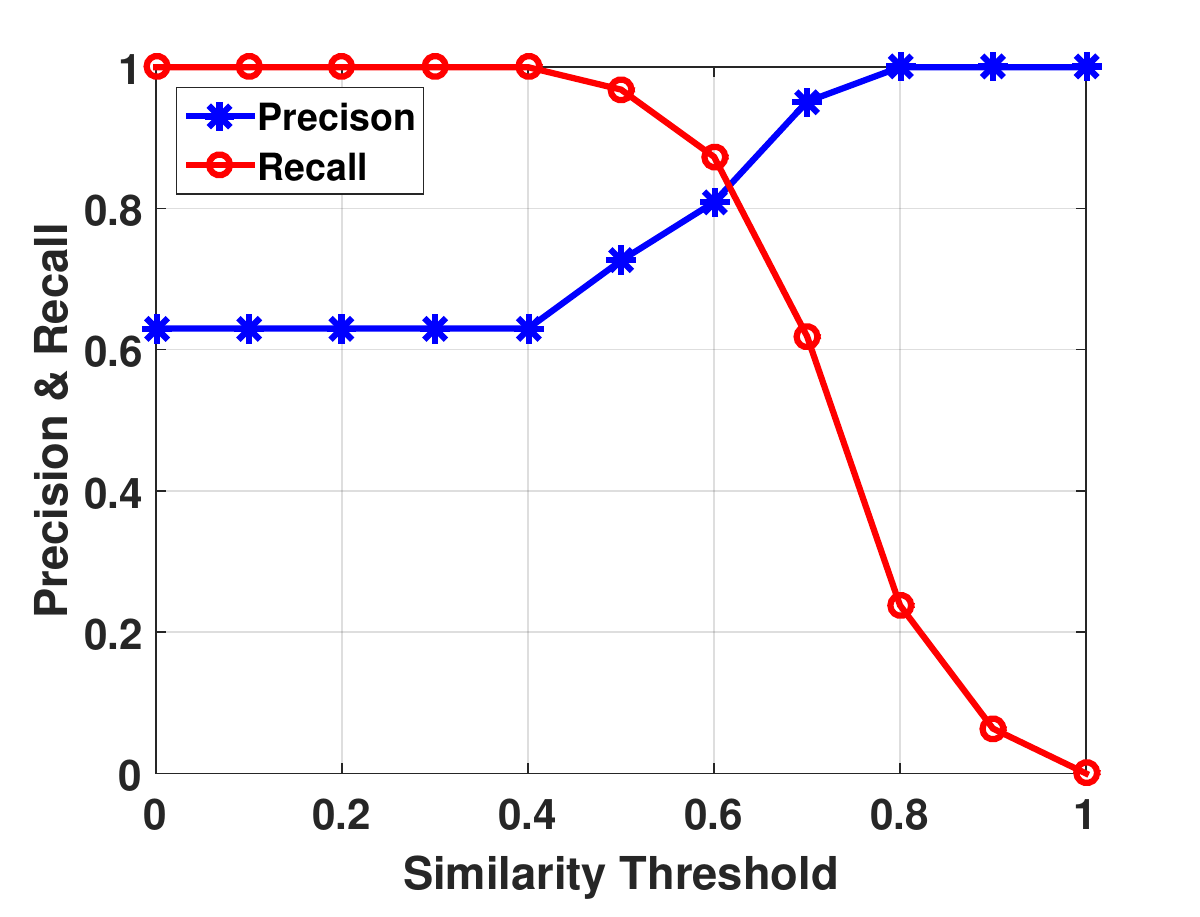}}
		\end{subfigure}\hfill
		\caption{Precision-recall curve for different values of similarity threshold for $\mathrm{D1}$ and for $\mathrm{Experiment~1}$ in which we use all the attributes for the profile matching.}
		\label{fig:recallall1}
	\end{minipage}%
	\hspace{0.5cm}
	\begin{minipage}{0.45\textwidth}
		\centering
		\begin{subfigure}[Global Attack]{\includegraphics[scale=0.19]{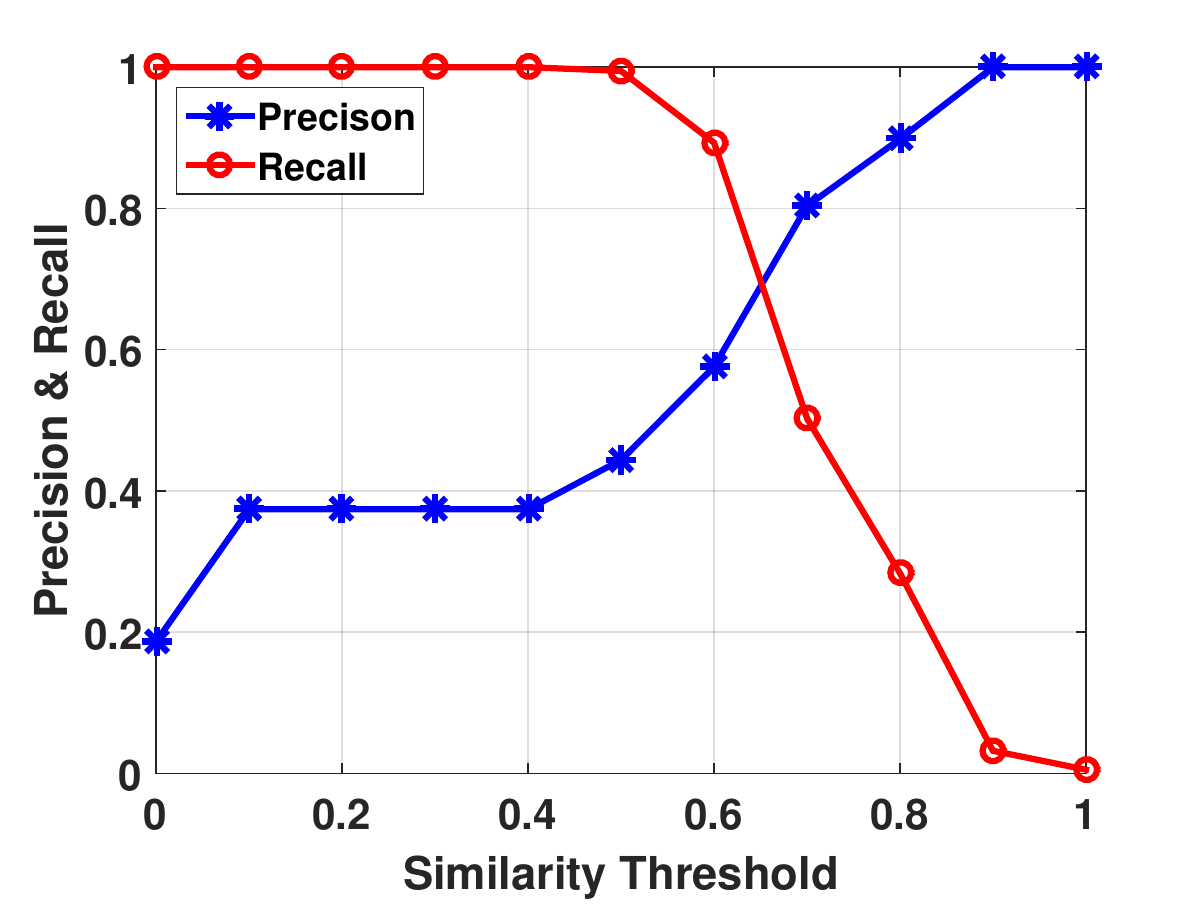}}
		\end{subfigure}\hfill
		\begin{subfigure}[Targeted Attack]{\includegraphics[scale=0.19]{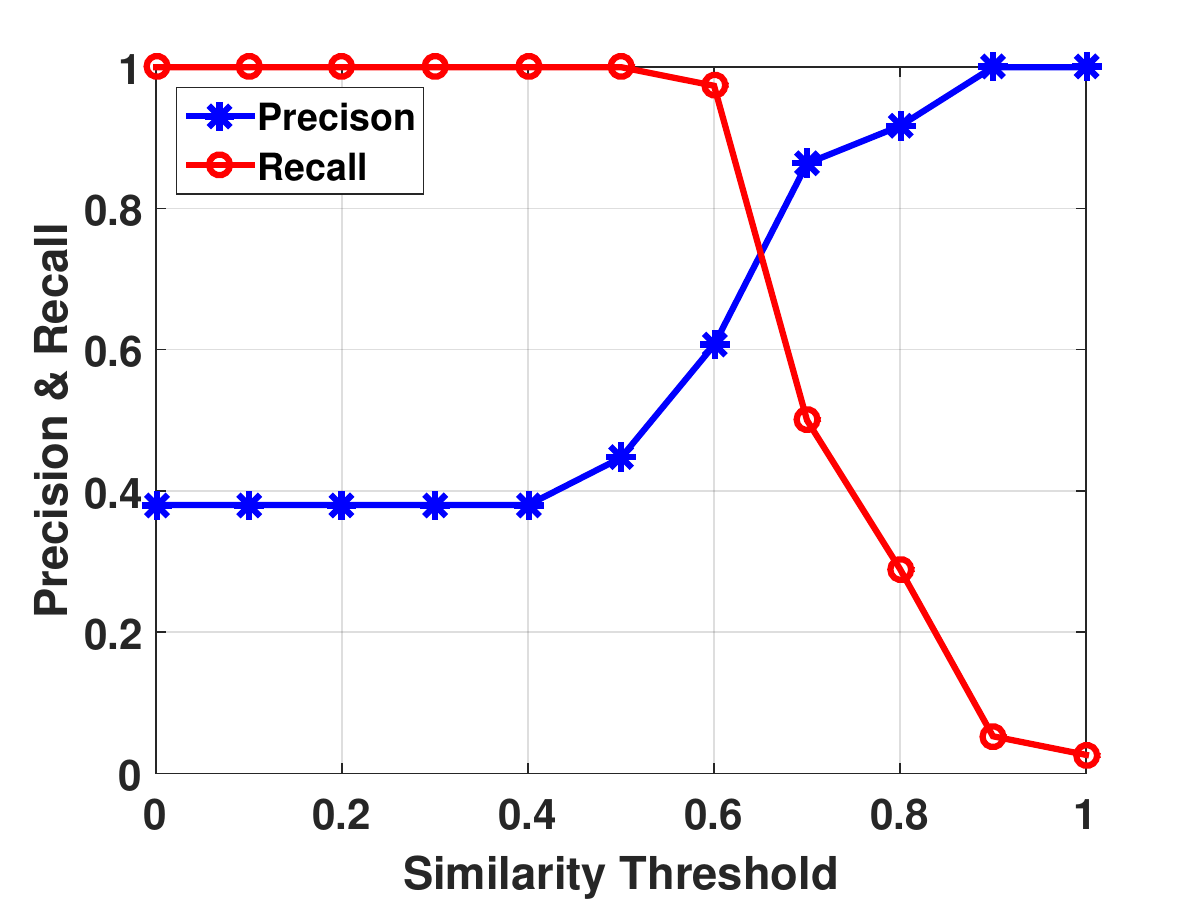}}
		\end{subfigure}\hfill
		\caption{Precision-recall curve for different values of similarity threshold for $\mathrm{D1}$ and for $\mathrm{Experiment~2}$ in which we only use the most identifying attributes for profile matching.}
		\label{fig:recallall21}
	\end{minipage}
	\begin{minipage}{0.45\textwidth}
		\centering
		\begin{subfigure}[Global Attack]{\includegraphics[scale=0.19]{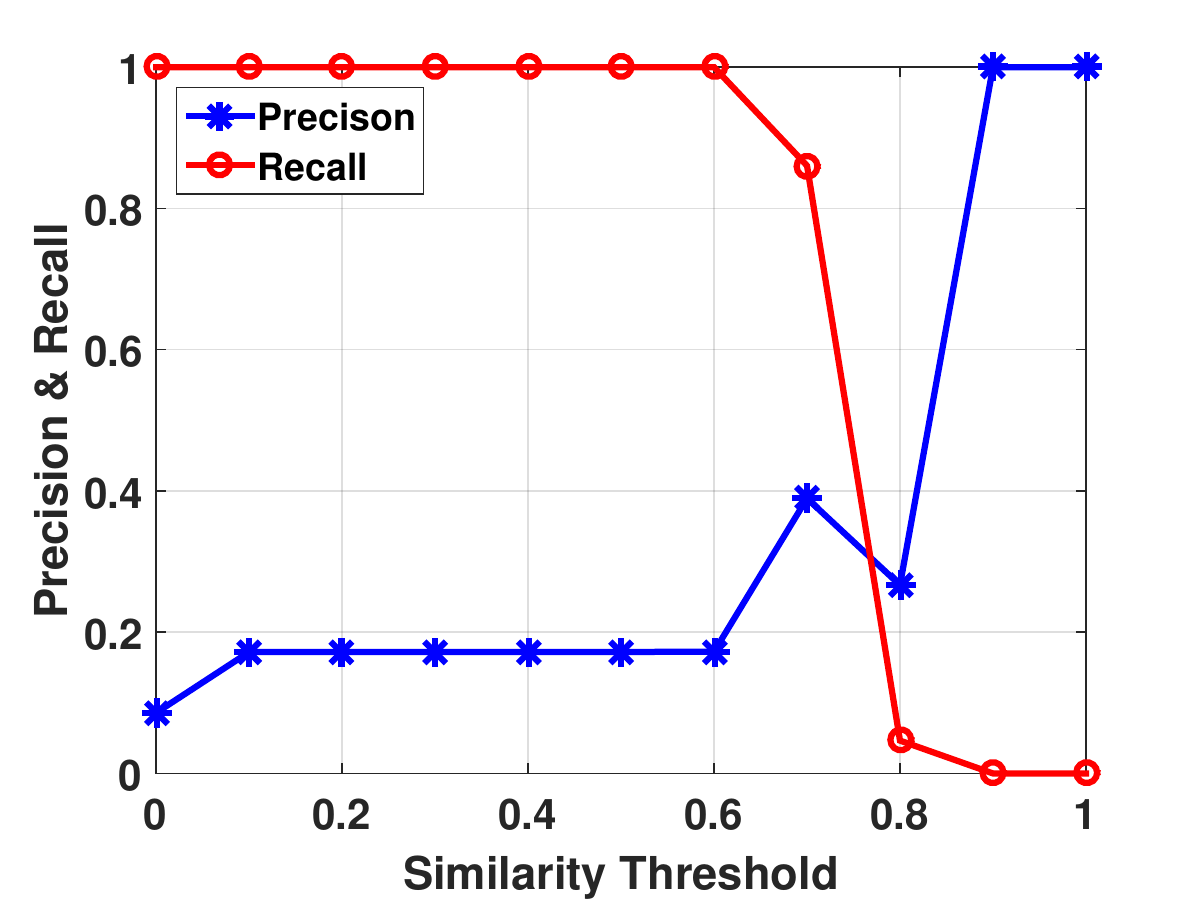}}
		\end{subfigure}\hfill
		\begin{subfigure}[Targeted Attack]{\includegraphics[scale=0.19]{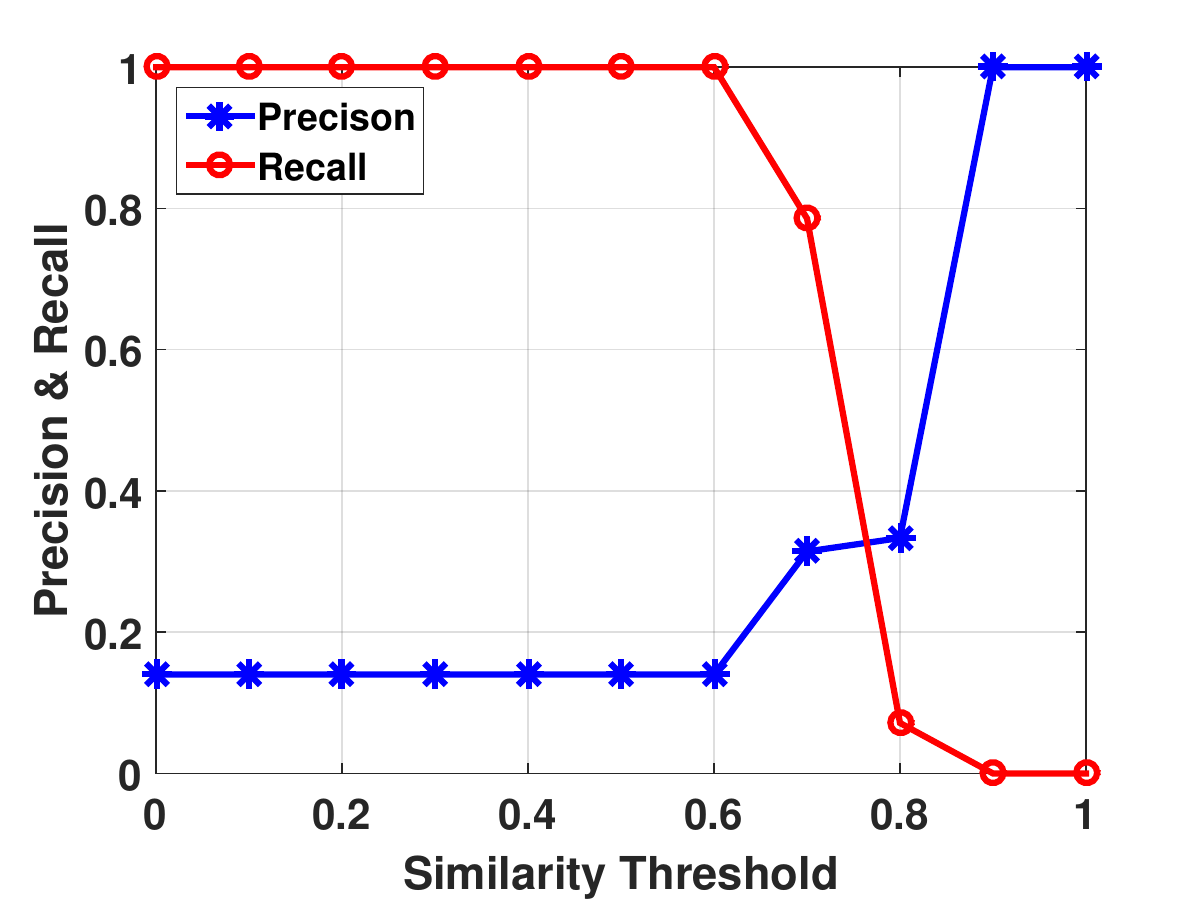}}
		\end{subfigure}\hfill
		\caption{Precision-recall curve for different values of similarity threshold for $\mathrm{D1}$ and for $\mathrm{Experiment~3}$ in which we use all the attributes except for the user name for profile matching.}
		\label{fig:recallall31}
	\end{minipage}
	\hspace{0.5cm}
	\begin{minipage}{0.45\textwidth}
		\centering
		\begin{subfigure}[Global Attack]{\includegraphics[scale=0.19]{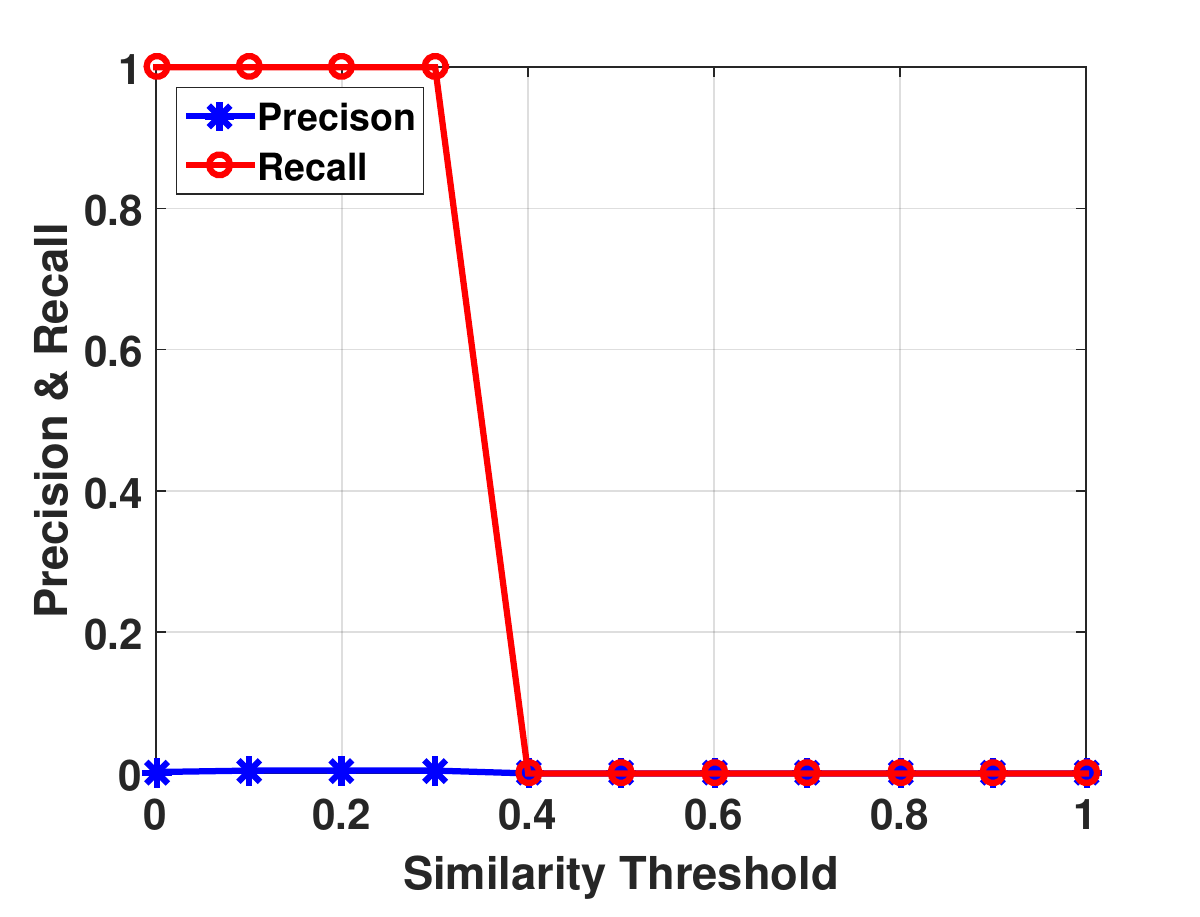}}
		\end{subfigure}\hfill
		\begin{subfigure}[Targeted Attack]{\includegraphics[scale=0.19]{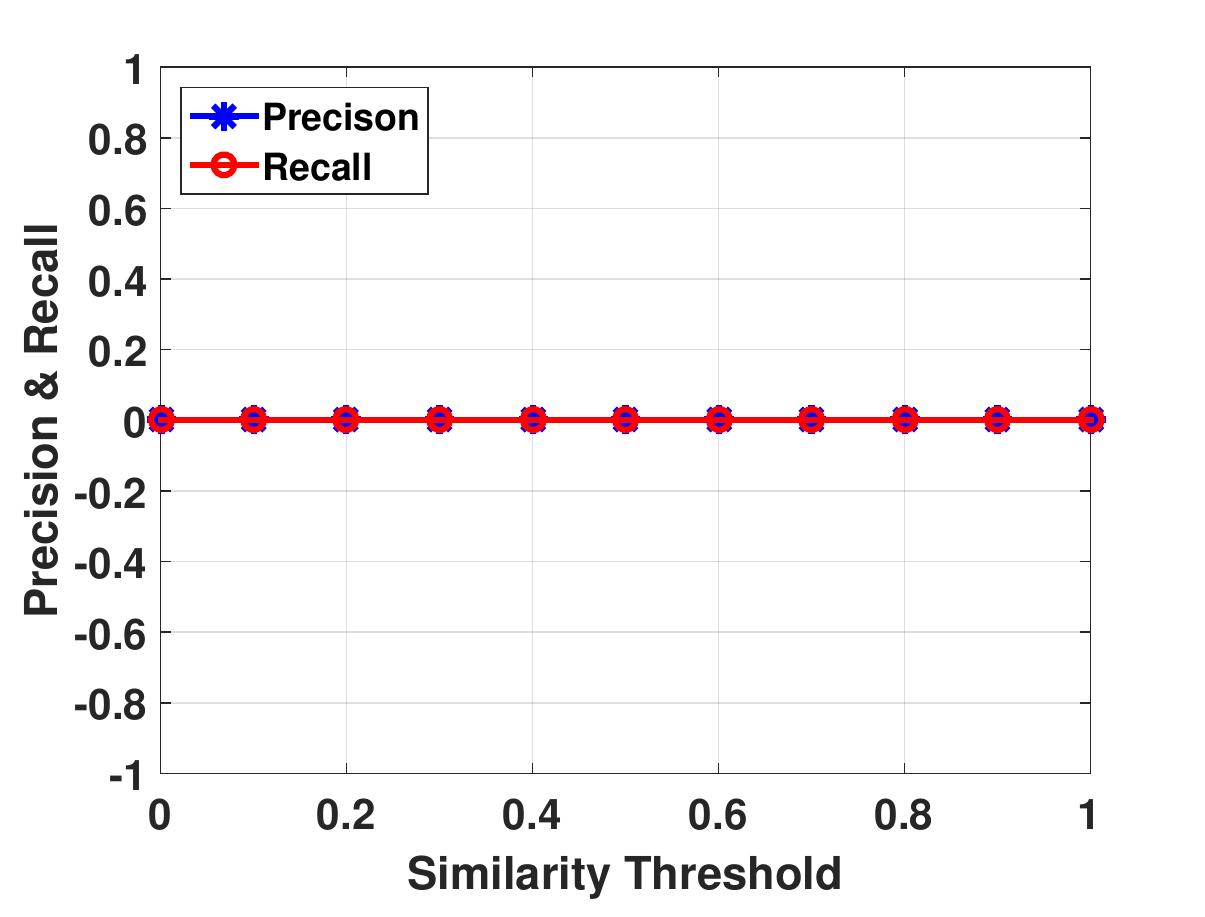}}
		\end{subfigure}\hfill
		\caption{Precision-recall curve for different values of similarity threshold for $\mathrm{D1}$ and for $\mathrm{Experiment~4}$ in which we use the weakest identifiers for profile matching.}
		\label{fig:recallall41}
	\end{minipage}
\end{figure*}

\begin{figure*}
	\centering
	\begin{minipage}{0.45\textwidth}
		\centering
		\begin{subfigure}[Global Attack]{\includegraphics[scale=0.19]{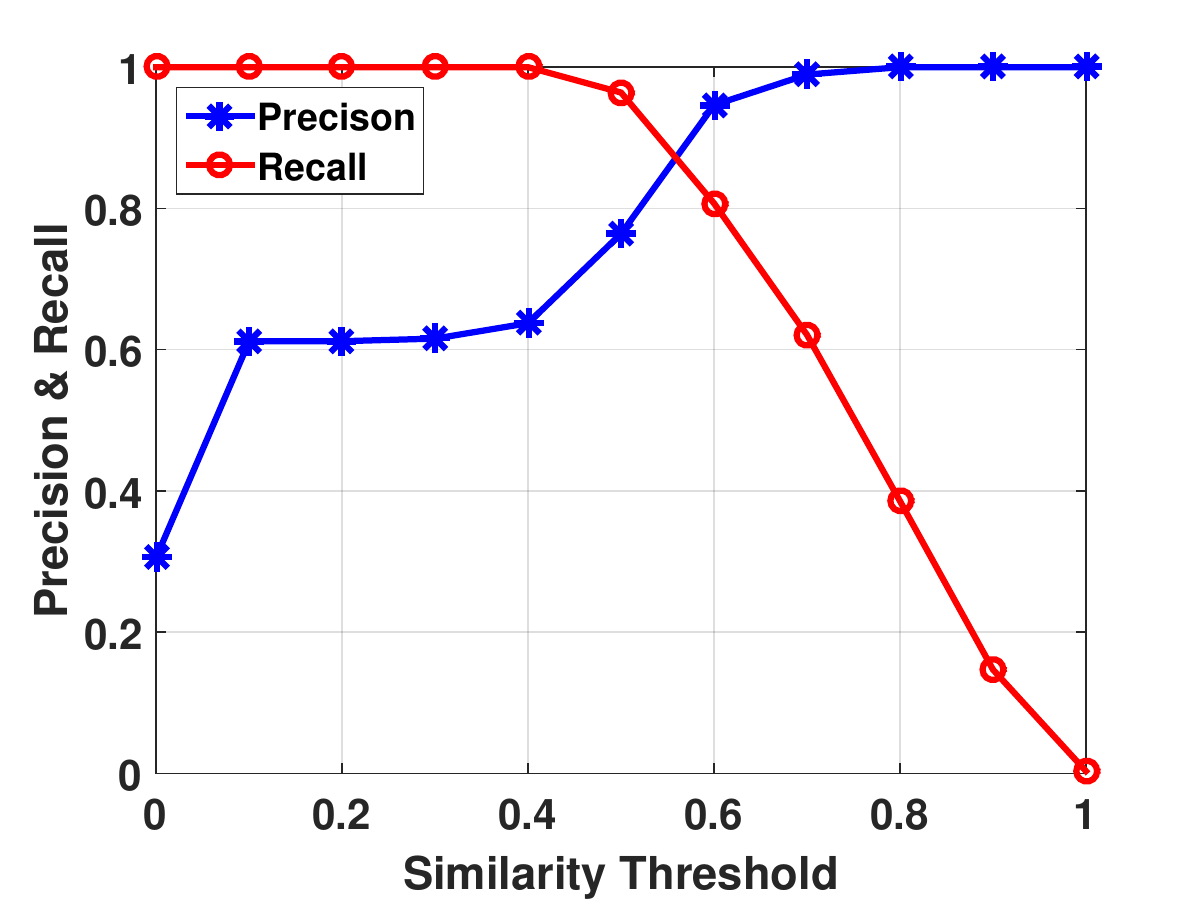}}
		\end{subfigure}\hfill
		\begin{subfigure}[Targeted Attack]{\includegraphics[scale=0.19]{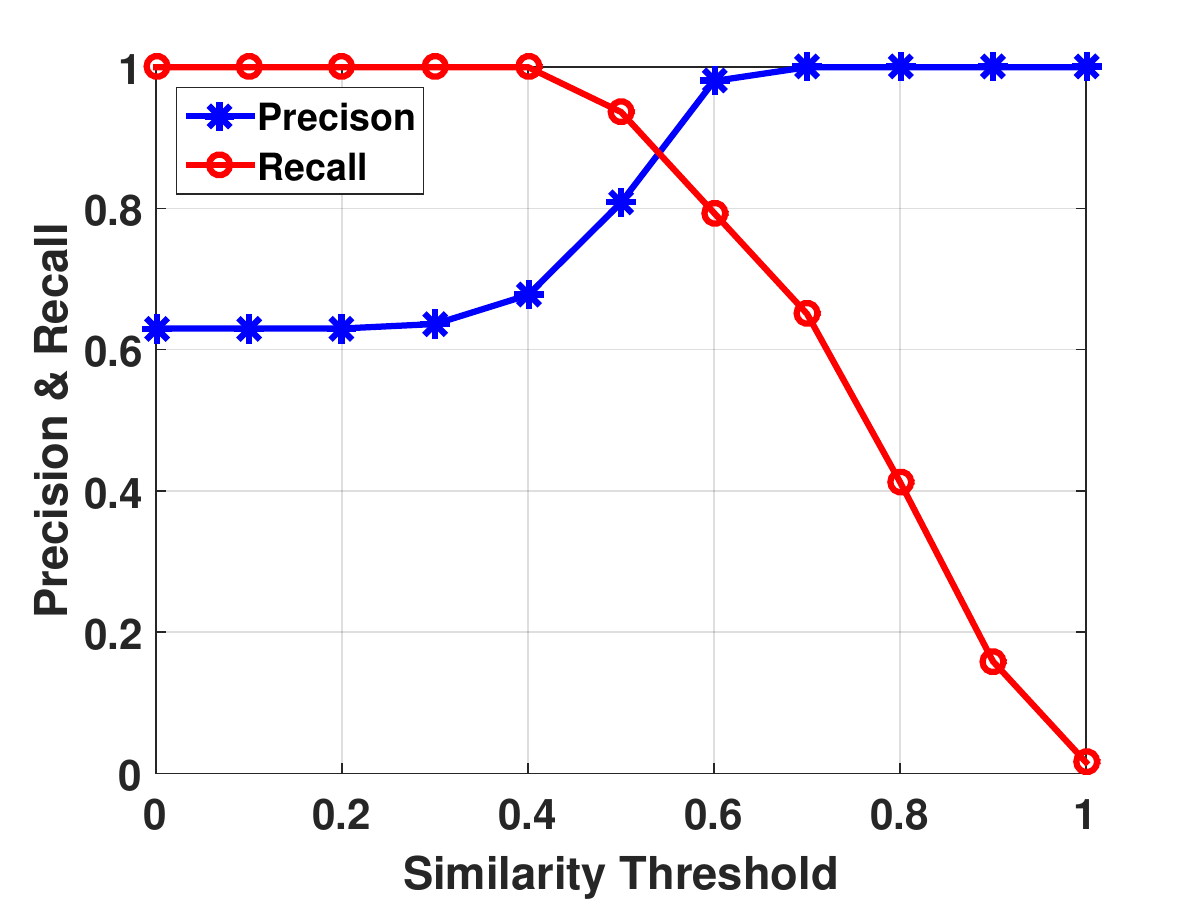}}
		\end{subfigure}\hfill
		\caption{Precision-recall curve for different values of similarity threshold for $\mathrm{D2}$ and for $\mathrm{Experiment~1}$ in which we use all the attributes for the profile matching.}
		\label{fig:recallall}
	\end{minipage}%
	\hspace{0.5cm}
	\begin{minipage}{0.45\textwidth}
		\centering
		\begin{subfigure}[Global Attack]{\includegraphics[scale=0.19]{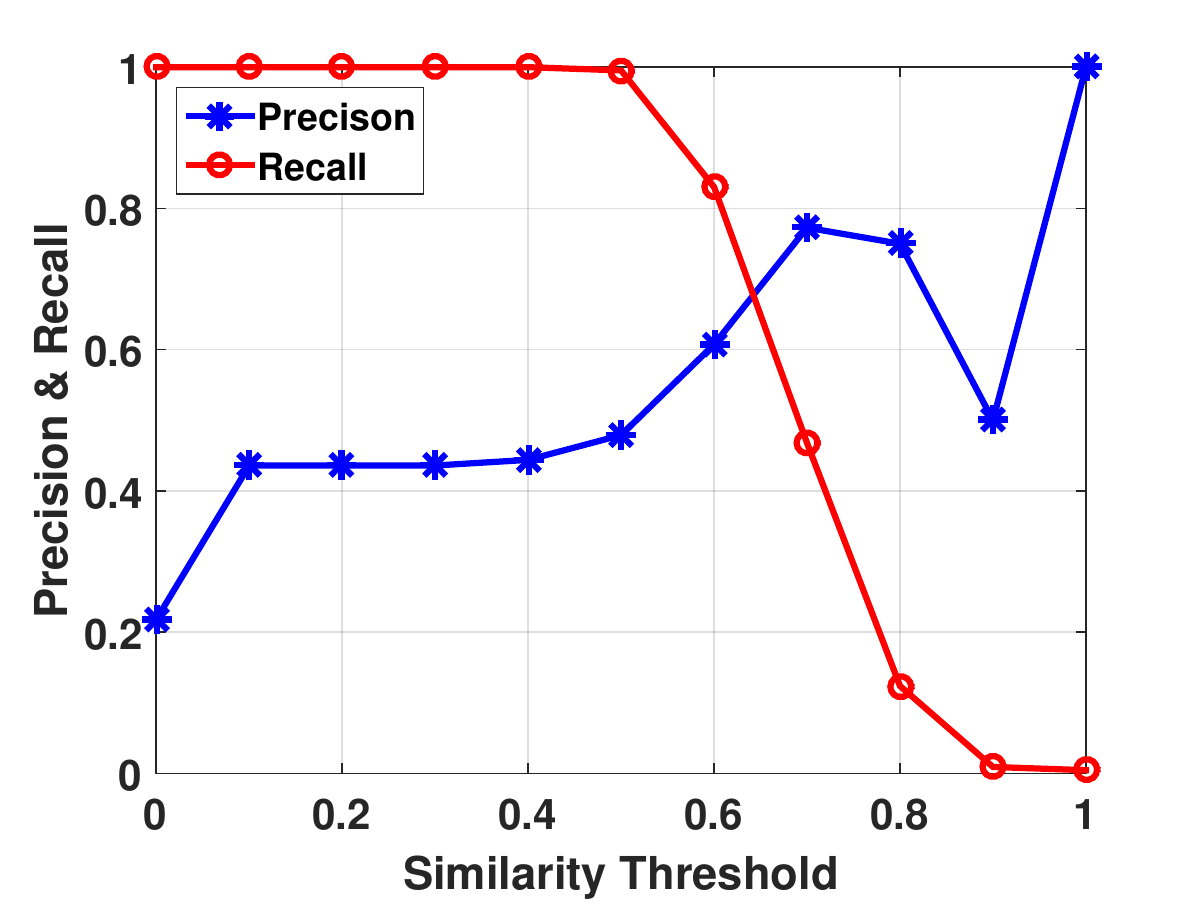}}
		\end{subfigure}\hfill
		\begin{subfigure}[Targeted Attack]{\includegraphics[scale=0.19]{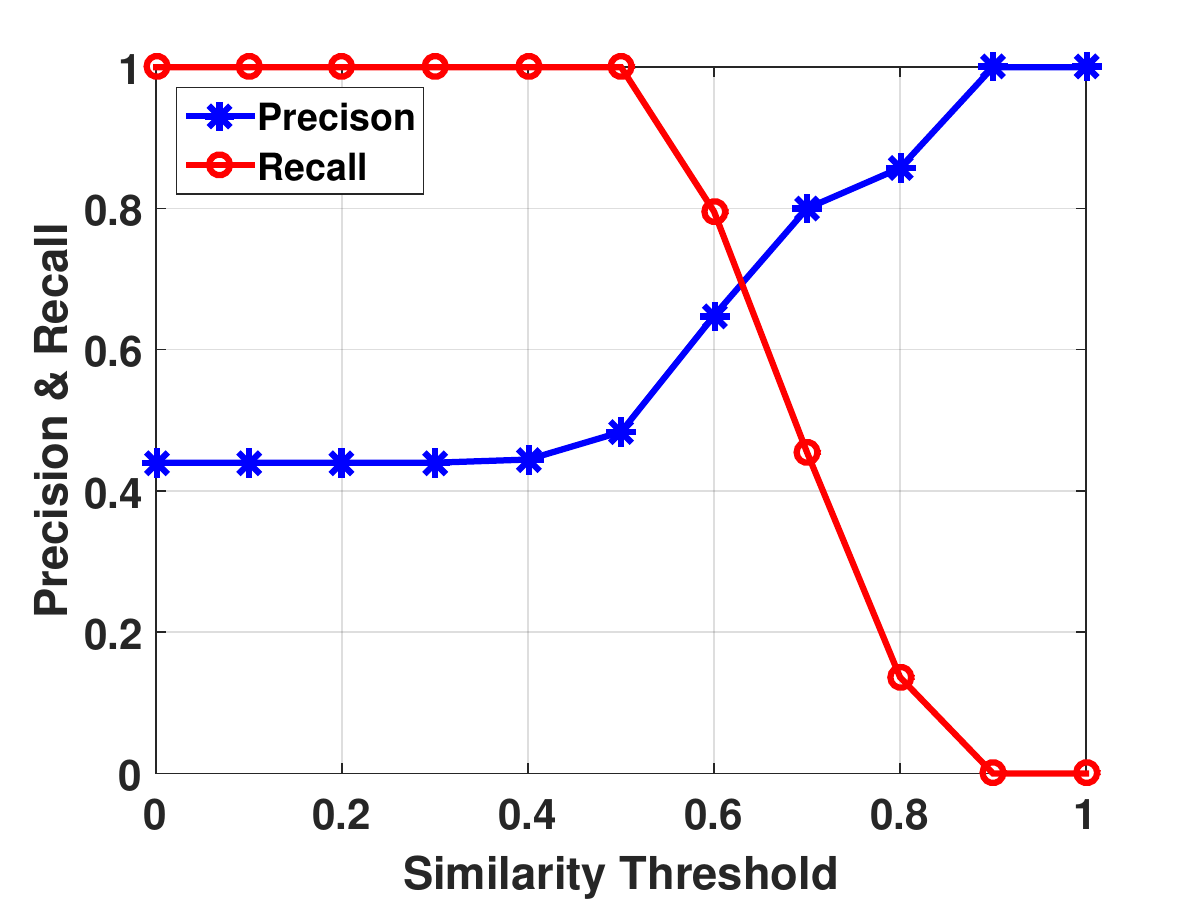}}
		\end{subfigure}\hfill
		\caption{Precision-recall curve for different values of similarity threshold for $\mathrm{D2}$ and for $\mathrm{Experiment~2}$ in which we only use the most identifying attributes for profile matching.}
		\label{fig:recallall2}
	\end{minipage}
	\begin{minipage}{0.45\textwidth}
		\centering
		\begin{subfigure}[Global Attack]{\includegraphics[scale=0.19]{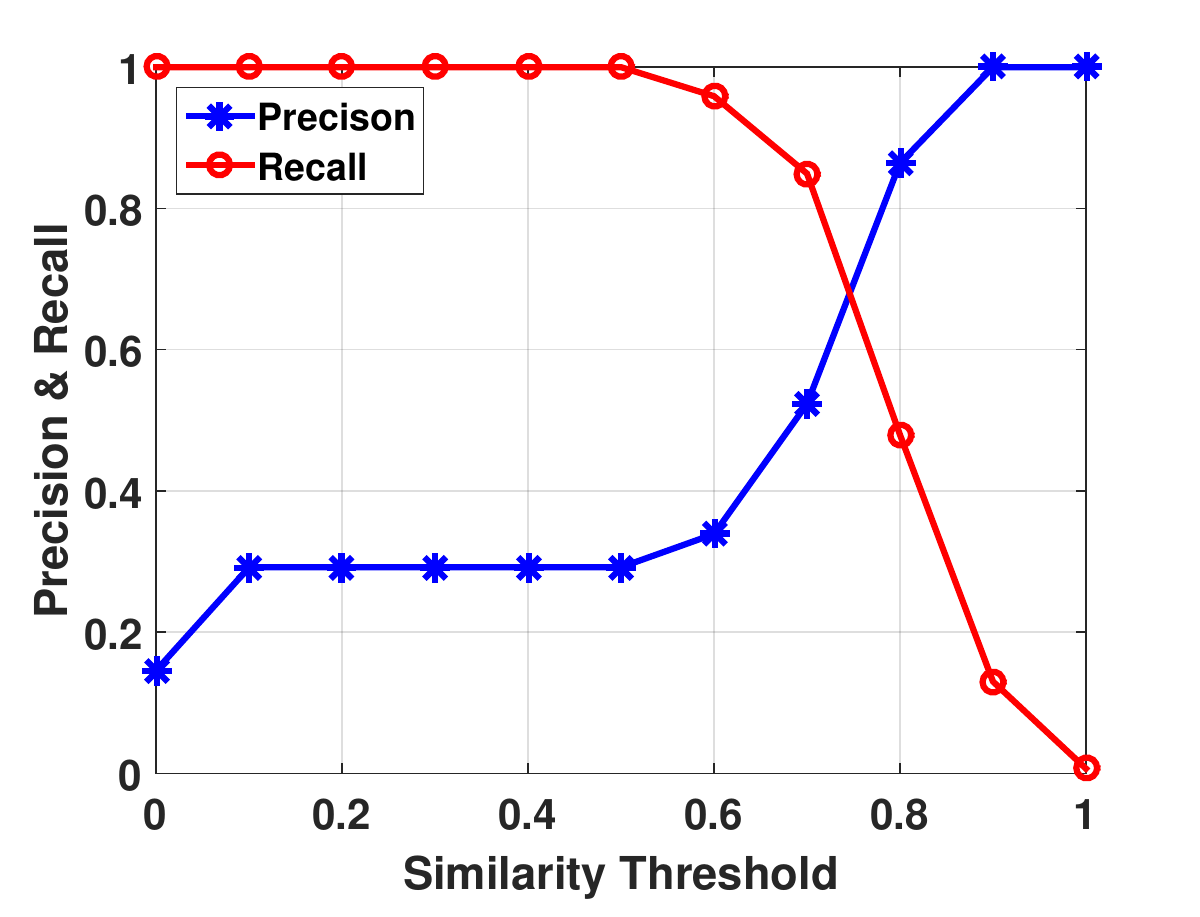}}
		\end{subfigure}\hfill
		\begin{subfigure}[Targeted Attack]{\includegraphics[scale=0.19]{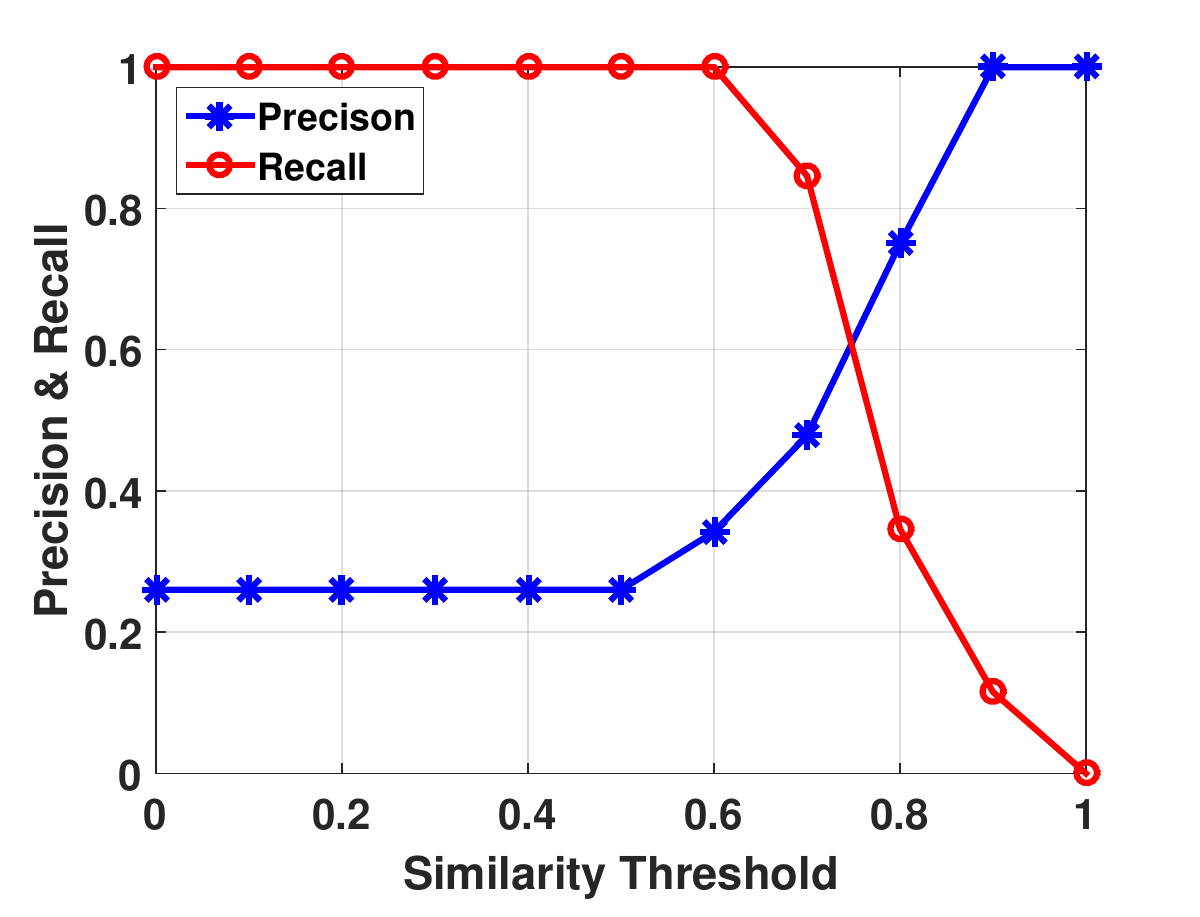}}
		\end{subfigure}\hfill
		\caption{Precision-recall curve for different values of similarity threshold for $\mathrm{D2}$ and for $\mathrm{Experiment~3}$ in which we use all the attributes except for the user name for profile matching.}
		\label{fig:recallall3}
	\end{minipage}
	\hspace{0.5cm}
	\begin{minipage}{0.45\textwidth}
		\centering
		\begin{subfigure}[Global Attack]{\includegraphics[scale=0.19]{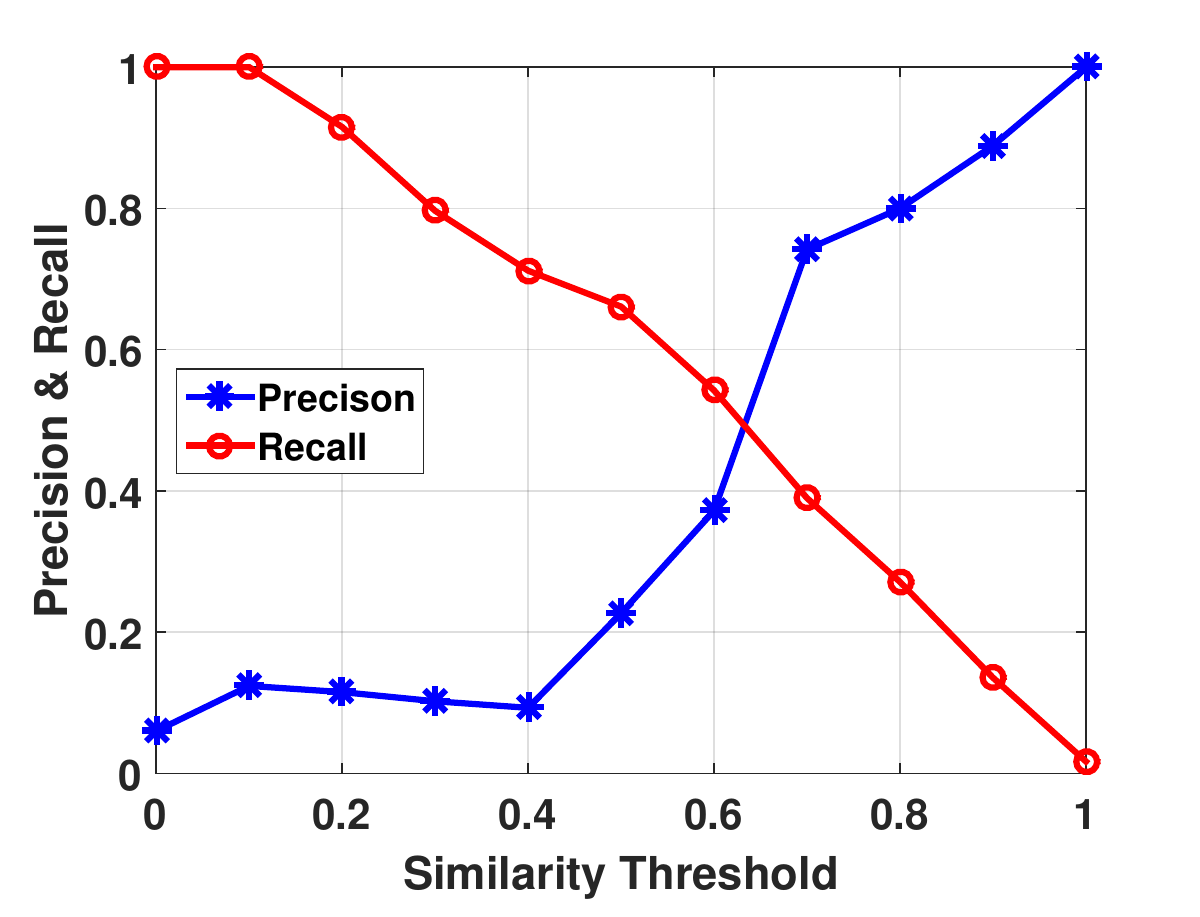}}
		\end{subfigure}\hfill
		\begin{subfigure}[Targeted Attack]{\includegraphics[scale=0.19]{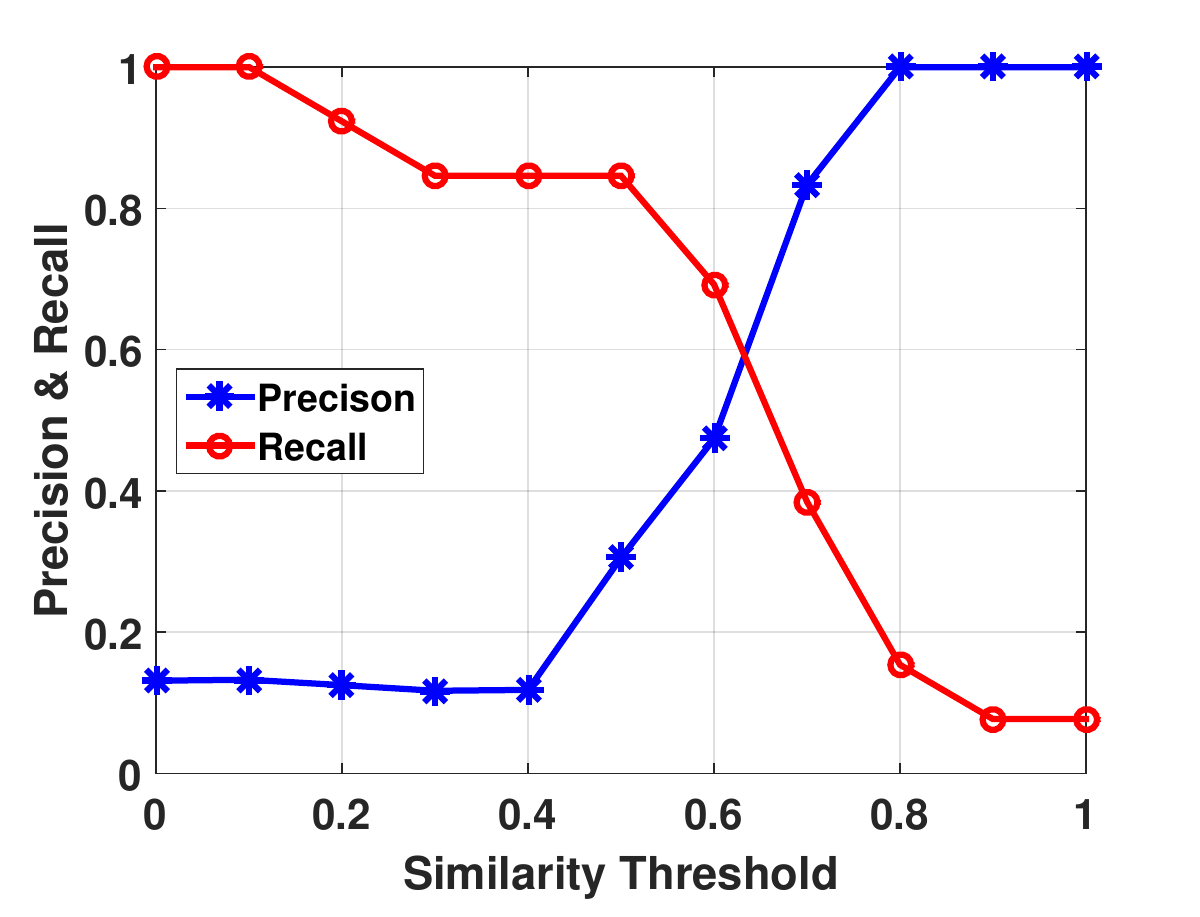}}
		\end{subfigure}\hfill
		\caption{Precision-recall curve for different values of similarity threshold for $\mathrm{D2}$ and for $\mathrm{Experiment~4}$ in which we use the weakest identifiers for profile matching.}
		\label{fig:recallall4}
	\end{minipage}
\end{figure*}

In real life, two OSNs do not contain exactly the same set of users. Thus, first, we evaluate the proposed attack by using a dataset that includes both coupled and uncoupled users. For the global attack, we tried to match all $N=1000$ profiles in $\mathrm{A_e}$ to $N=1000$ profiles in $\mathrm{T_e}$. Among these profiles, $500$ of them are coupled profiles, and hence the goal is to make sure that these $500$ users are matched with high confidence. In targeted attack, we set the number of target individuals to $100$ from $T$. These $100$ coupled profiles for the targeted attack are randomly picked among 500 coupled pairs in the test dataset. We run the targeted attack for $10$ times and get the average of the results. We ran all four experiments (introduced in Section~\ref{sec:learning}) for these settings. For the clarity of the presentation, in the following, we refer the Twitter - Foursquare as $\mathrm{D1}$ and Google+ - Twitter as $\mathrm{D2}$. For all experiments, we report the precision and recall values for the similarity threshold at which the precision and recall curves (almost) intersect. In Table~\ref{table:results}, we present the results obtained for the linear regression model and in Table~\ref{table:results_svm}, the results for the SVM model are shown. 
The variation of the precision and recall values with the similarity threshold for both $\mathrm{D1}$ and $\mathrm{D2}$ and for all four experiments (using the linear model regression ) are shown in Figures~\ref{fig:recallall1} -~\ref{fig:recallall4}.
In general, we observed that the precision, recall, and success rate of the linear regression model are higher compared to the SVM model. 
\begin{table*}[t]
	\centering
	\resizebox{\textwidth}{!}{
		\begin{tabular}{l|c|c|c|c|c|c|c|c|c|c|c|c|}
			\cline{2-13}
			& \multicolumn{6}{c|}{Dataset~1 (Twitter - Foursquare)} & \multicolumn{6}{c|}{Dataset~2 (Google+ - Twitter)}\\
			\cline{2-13}
			& \multicolumn{3}{c|}{Global Attack} & \multicolumn{3}{c|}{Targeted Attack} & \multicolumn{3}{c|}{Global Attack} & \multicolumn{3}{c|}{Targeted Attack}\\
			\cline{2-13}
			& Precision & Recall & Success rate & Precision & Recall & Success rate & Precision & Recall & Success rate & Precision & Recall & Success rate\\
			\hline
			\multicolumn{1}{|l|}{$\mathrm{Experiment~1}$ (with all attributes)} & $0.79$ & $0.79$ & $58.6\%$ & $0.85$ & $0.85$ & $63\%$ & $0.88$ & $0.89$ & $62\%$ & $0.88$ & $0.89$ & $63\%$\\
			\hline
			\multicolumn{1}{|l|}{$\mathrm{Experiment~2}$ (with the most identifying attributes)} & $0.71$ & $0.73$ & $37.5\%$ & $0.73$ & $0.73$ & $38\%$ & $0.68$ & $0.68$ & $43.6\%$ & $0.69$ & $0.69$ & $44\%$\\
			\hline
			\multicolumn{1}{|l|}{$\mathrm{Experiment~3}$ (excluding the user name)} & $0.2$ & $0.18$ & $17.2\%$ & $0.2$ & $0.19$ & $14\%$ & $0.63$ & $0.60$ & $29.2\%$ & $0.60$ & $0.62$ & $26\%$\\
			\hline
			\multicolumn{1}{|l|}{$\mathrm{Experiment~4}$ (with the weakest identifiers)} & $0.004$ & $0.004$ & $0.4\%$ & $\sim0$ & $\sim0$ & $0\%$ & $0.45$ & $0.46$ &$12\%$ &$0.43$ & $0.43$ & $13\%$\\
			\hline
		\end{tabular}
	}
	\caption{Results of the profile matching attack (both targeted and global) with both coupled and uncoupled profiles by using linear regression as the machine learning technique. For all experiments, we report the precision and recall values for the similarity threshold at which the precision and recall curves (almost) intersect.}
	\label{table:results}
\end{table*}

\begin{table*}[t]
	\centering
	\resizebox{\textwidth}{!}{
		\begin{tabular}{l|c|c|c|c|c|c|c|c|c|c|c|c|}
			\cline{2-13}
			& \multicolumn{6}{c|}{Dataset~1 (Twitter - Foursquare)} & \multicolumn{6}{c|}{Dataset~2 (Google+ - Twitter)}\\
			\cline{2-13}
			& \multicolumn{3}{c|}{Global Attack} & \multicolumn{3}{c|}{Targeted Attack} & \multicolumn{3}{c|}{Global Attack} & \multicolumn{3}{c|}{Targeted Attack}\\
			\cline{2-13}
			& Precision & Recall & Success rate & Precision & Recall & Success rate & Precision & Recall & Success rate & Precision & Recall & Success rate\\
			\hline
			\multicolumn{1}{|l|}{$\mathrm{Experiment~1}$ (with all attributes)} & $0.60$ & $0.59$ & $41\%$ & $0.72$ & $0.73$ & $54.5\%$ & $0.90$ & $0.89$ & $50\%$ & $0.87$ & $0.86$ & $62.6\%$\\
			\hline
			\multicolumn{1}{|l|}{$\mathrm{Experiment~2}$ (with the most identifying attributes)} & $0.68$ & $0.67$ & $38.6\%$ & $0.42$ & $0.41$ & $40\%$ & $0.42$ & $0.42$ & $41\%$ & $0.41$ & $0.43$ & $46\%$\\
			\hline
			\multicolumn{1}{|l|}{$\mathrm{Experiment~3}$ (excluding the user name)} & $0.08$ & $0.07$ & $11\%$ & $\sim0$ & $\sim0$ & $\sim0\%$ & $0.32$ & $0.33$ & $18\%$ & $0.21$ & $0.27$ & $13\%$\\
			\hline
			\multicolumn{1}{|l|}{$\mathrm{Experiment~4}$ (with the weakest identifiers)} & $\sim0$ & $\sim0$ & $\sim0\%$ & $\sim0$ & $\sim0$ & $\sim0\%$ & $0.33$ & $0.32$ & $13.4\%$ & $0.47$ & $0.45$ & $14\%$\\
			\hline
		\end{tabular}
	}
	\caption{Results of the profile matching attack (both targeted and global) with both coupled and uncoupled profiles by using SVM as the machine learning technique. Precision and recall values are computed with the similarity threshold at which the precision and recall curves (almost) intersect.}
	\label{table:results_svm}
\end{table*}

\begin{figure}
	\begin{minipage}{0.50\textwidth}
		\centering
		\begin{subfigure}[Database~1]{\includegraphics[scale=0.23]{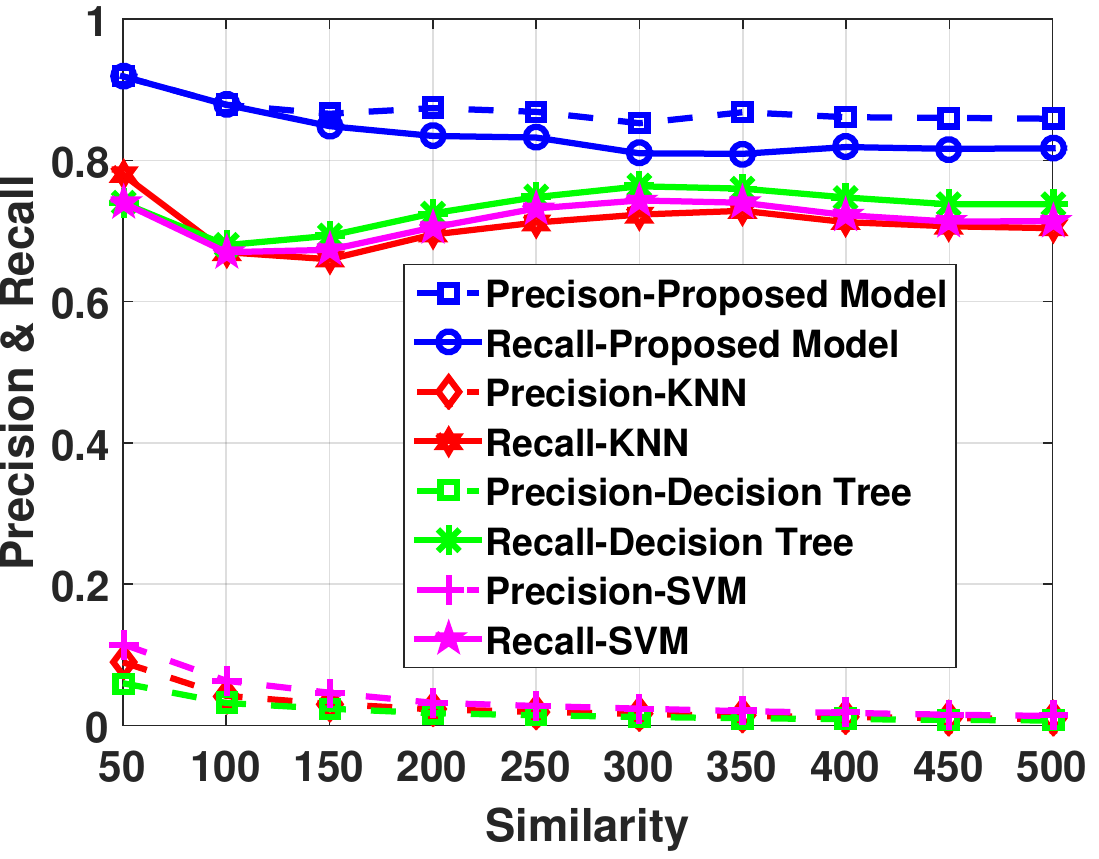}}
		\end{subfigure}\hfill
		\begin{subfigure}[Database~2]{\includegraphics[scale=0.23]{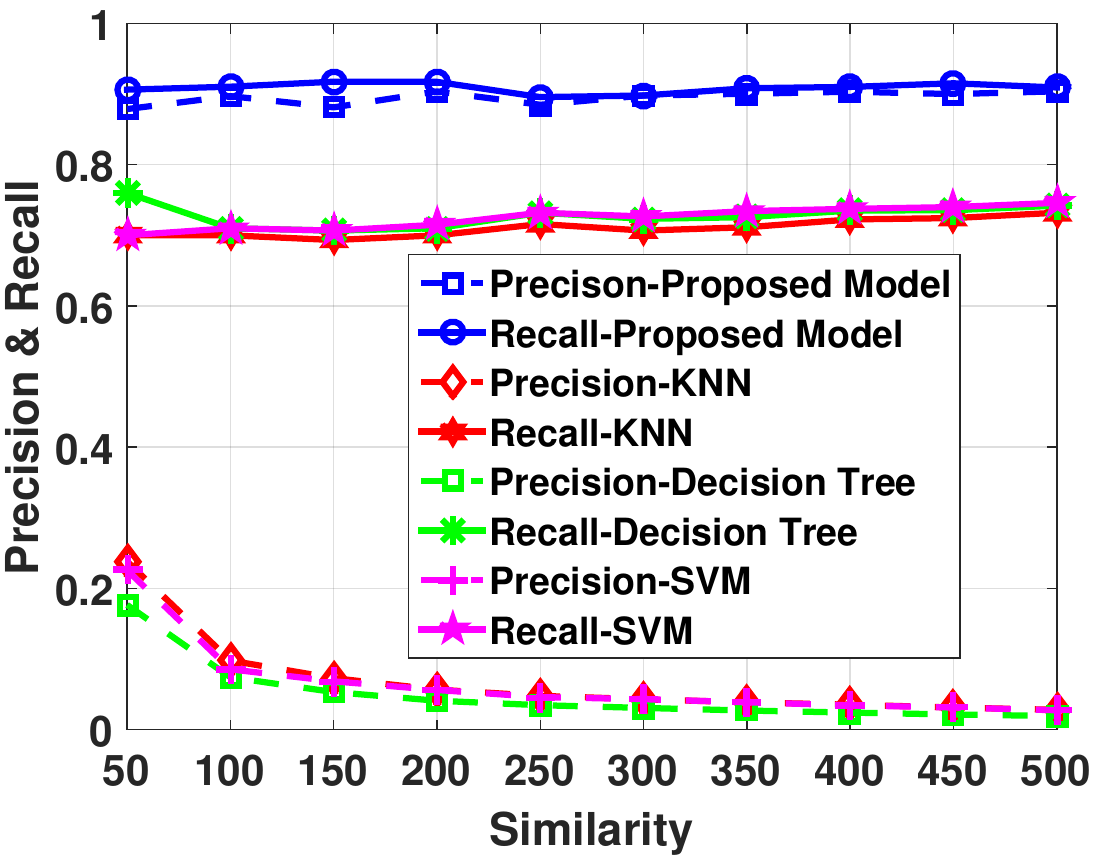}}
		\end{subfigure}\hfill
		\caption{The effect of database size to the precision/recall for the global attack in $\mathrm{Experiment~1}$ with only coupled profiles.}
		\label{fig:coupled_attack}
	\end{minipage}
\end{figure}

In $\mathrm{Experiment~1}$ (in which we use all the considered attributes), for the global attack, we obtained a precision value of around $0.80$ (for $\mathrm{D1}$) and $0.90$ (for $\mathrm{D2}$) for a similarity threshold of $0.6$. This means that if our proposed attack returns a similarity value that is above $0.6$ for a given profile pair, we can say that the corresponding profiles belong to same individual with a high confidence. Also, overall, we could correctly match $293$ coupled profiles in $\mathrm{D1}$ (with a success rate of $58.6\%$) and $306$ coupled profiles in $\mathrm{D2}$ (with a success rate of $62\%$) out of $500$ in global attack. Furthermore, in targeted attack, we obtained a precision value of $0.85$ for $\mathrm{D1}$ and $0.88$ for $\mathrm{D2}$ (for a similarity threshold of 0.6) and overall, we were able to correctly match 62 profiles in $\mathrm{D1}$ and 63 profiles in $\mathrm{D2}$ (out of 100). Using the same test dataset, we obtained a precision that is close to zero by using the baseline approach (both by using KNN and decision trees and for both datasets). This shows that the proposed framework significantly improves the baseline approach while it provides comparable recall value compared to these machine learning techniques (this is further discussed in Figure~\ref{fig:coupled_attack}).

In $\mathrm{Experiment~2}$ (in which we use the four most identifying attributes based on the weights in Table~\ref{table:weights1}), we obtained a success rate of $37.5\%$ (for $\mathrm{D1}$) and $43.6\%$ (for $\mathrm{D2}$) for the global attack with a precision value of around $0.71$ and $0.68$ for $\mathrm{D1}$ and $\mathrm{D2}$, respectively. Also, we could match 38 profiles (in $\mathrm{D1}$) and 44 profiles (in $\mathrm{D2}$) out of 100 in the targeted attack. In $\mathrm{Experiment~3}$ (in which we use all the attributes but the user name), for the global attack, we obtained a precision value of around $0.25$ (for $\mathrm{D1}$) and $0.63$ (for $\mathrm{D2}$)  and an overall success rate of $14\%$ and $26\%$ for $\mathrm{D1}$ and $\mathrm{D2}$, respectively. Finally, in $\mathrm{Experiment~4}$ (in which we use the weakest attributes), for the global attack, we obtained a precision value of almost $0$ (for $\mathrm{D1}$) and $0.45$ (for $\mathrm{D2}$)  and an overall success rate of $12\%$ for $\mathrm{D2}$. Furthermore, in $\mathrm{D2}$, we correctly matched 13 profiles out of 100 in the targeted attack. We further comment on these results in the next section. Overall, the results show that publicly sharing identifying attributes significantly helps the profile matching attack. Furthermore, we show that even the weakest identifiers may cause profile matching between the OSN users for some cases.


Next, only using the $500$ coupled profiles in our test dataset, first we ran all four experiments (introduced in Section~\ref{sec:learning}) as before, and then we studied the effects of database size to the profile matching attack. Thus, for the global attack, we tried to match all $N=500$ profiles in $\mathrm{A_e}$ to $N=500$ profiles in $\mathrm{T_e}$ (all these are coupled profiles this time) and in targeted attack, we set the number of target individuals to $100$ from $T$ as before. We show the success rate (i.e., fraction of the correctly matched profiles as a result of the attack) and precision/recall values we get from each experiment for the linear regression model in Table~\ref{table:coupled_attack}. The results for the SVM model are also shown in Table~\ref{table:coupled_attack_svm}. As before, in general, we obtained more accurate results for the linear regression model compared to the SVM model. The precision and recall values reported in both tables are obtained when we set the similarity threshold to the value at which the precision and recall curves (almost) intersect. In practice, the attacker can pick the similarity threshold based on the set of attributes being used for the attack. In general, we observed that all precision, recall, and success rate values we obtained for this scenario are higher than the ones reported for the previous attack scenario (in Table~\ref{table:results}).
\begin{table*}[t]
	\centering
	\resizebox{\textwidth}{!}{
		\begin{tabular}{l|c|c|c|c|c|c|c|c|c|c|c|c|}
			\cline{2-13}
			& \multicolumn{6}{c|}{Dataset~1 (Twitter - Foursquare)} & \multicolumn{6}{c|}{Dataset~2 (Google+ - Twitter)}\\
			\cline{2-13}
			& \multicolumn{3}{c|}{Global Attack} & \multicolumn{3}{c|}{Targeted Attack} & \multicolumn{3}{c|}{Global Attack} & \multicolumn{3}{c|}{Targeted Attack}\\
			\cline{2-13}
			& Precision & Recall & Success rate & Precision & Recall & Success rate & Precision & Recall & Success rate & Precision & Recall & Success rate\\
			\hline
			\multicolumn{1}{|l|}{$\mathrm{Experiment~1}$ (with all attributes)} & $0.82$ & $0.83$ & $65.6\%$ & $0.87$ & $0.87$ & $66\%$ & $0.90$ & $0.90$ & $66.2\%$ & $0.92$ & $0.92$ & $72\%$\\
			\hline
			\multicolumn{1}{|l|}{$\mathrm{Experiment~2}$ (with the most identifying attributes)} & $0.71$ & $0.73$ & $43\%$ & $0.72$ & $0.73$ & $45\%$ & $0.76$ & $0.76$ & $54\%$ & $0.74$ & $0.74$ & $56\%$\\
			\hline
			\multicolumn{1}{|l|}{$\mathrm{Experiment~3}$ (excluding the user name)} & $0.26$ & $0.20$ & $19\%$ & $0.3$ & $0.28$ & $16\%$ & $0.60$ & $0.61$ & $36\%$ & $0.60$ & $0.60$ & $38\%$\\
			\hline
			\multicolumn{1}{|l|}{$\mathrm{Experiment~4}$ (with the weakest identifiers)} & $\sim0$ & $\sim0$ & $0.4\%$ & $\sim0$ & $\sim0$ & $1\%$ & $0.71$ & $0.69$ & $12.8\%$ & $0.66$ & $0.66$ & $13\%$\\
			\hline
		\end{tabular}
	}
	\caption{Results of the profile matching attack (both targeted and global) with only coupled profiles by using linear regression as the machine learning technique. Precision and recall values are computed with the similarity threshold at which the precision and recall curves (almost) intersect.}
	\label{table:coupled_attack}
\end{table*}

\begin{table*}[t]
	\centering
	\resizebox{\textwidth}{!}{
		\begin{tabular}{l|c|c|c|c|c|c|c|c|c|c|c|c|}
			\cline{2-13}
			& \multicolumn{6}{c|}{Dataset~1 (Twitter - Foursquare)} & \multicolumn{6}{c|}{Dataset~2 (Google+ - Twitter)}\\
			\cline{2-13}
			& \multicolumn{3}{c|}{Global Attack} & \multicolumn{3}{c|}{Targeted Attack} & \multicolumn{3}{c|}{Global Attack} & \multicolumn{3}{c|}{Targeted Attack}\\
			\cline{2-13}
			& Precision & Recall & Success rate & Precision & Recall & Success rate & Precision & Recall & Success rate & Precision & Recall & Success rate\\
			\hline
			\multicolumn{1}{|l|}{$\mathrm{Experiment~1}$ (with all attributes)} & $0.66$ & $0.67$ & $50.4\%$ & $0.76$ & $0.75$ & $52\%$ & $0.91$ & $0.92$ & $50.1\%$ & $0.91$ & $0.92$ & $53\%$\\
			\hline
			\multicolumn{1}{|l|}{$\mathrm{Experiment~2}$ (with the most identifying attributes)} & $0.67$ & $0.68$ & $42\%$ & $0.75$ & $0.76$ & $46\%$ & $0.71$ & $0.69$ & $50\%$ & $0.51$ & $0.50$ & $50\%$\\
			\hline
			\multicolumn{1}{|l|}{$\mathrm{Experiment~3}$ (excluding the user name)} & $0.08$ & $0.04$ & $12.8\%$ & $0.08$ & $0.01$ & $12\%$ & $0.30$ & $0.31$ & $24\%$ & $0.33$ & $0.3$ & $21\%$\\
			\hline
			\multicolumn{1}{|l|}{$\mathrm{Experiment~4}$ (with the weakest identifiers)} & $\sim0$ & $\sim0$ & $\sim0\%$ & $0.01$ & $0.02$ & $1\%$ & $0.21$ & $0.22$ & $11\%$ & $0.16$ & $0.19$ & $12\%$\\
			\hline
		\end{tabular}
	}
	\caption{Results of the profile matching attack (both targeted and global) with only coupled profiles by using SVM as the machine learning technique. Precision and recall values are computed with the similarity threshold at which the precision and recall curves (almost) intersect.}
	\label{table:coupled_attack_svm}
\end{table*}

Finally, in Figure~\ref{fig:coupled_attack}, we show the precision/recall of the proposed attack for $\mathrm{Experiment~1}$ as a function of the database size for the global attack and for the linear regression model. For the proposed attack, we report the precision and recall value for the similarity threshold at which the precision and recall curves almost intersect (as before). In the same figure, we also compare the proposed attack with the baseline approach in which we use KNN, decision trees and SVM for the profile matching as discussed in Section~\ref{sec:baseline}. We observed that the precision/recall of the proposed attack does not decrease with increasing database size, which shows the scalability of our proposed attack. We also observed that the proposed attack notably provides significantly higher precision values compared to the baseline approach.

\subsection{Discussion}\label{sec:discussion}

In general, for all experiments, we observed that linear regression provides better results compared to the SVM model. In terms of the variation of the results obtained for different databases,  we have the following observations:
\begin{itemize}
	\item Precision, recall, and success rate obtained from $\mathrm{D2}$ are higher compared to $\mathrm{D1}$ as users share more complete and informative information in Google+ compared to Foursquare. In particular, $\mathrm{Experiment~3}$ and $\mathrm{Experiment~4}$ show that Google+ profiles provide more complete information in terms of freetext sharings, activity patterns, and interests of the users.
	\item In $\mathrm{D1}$, the weight for the activity pattern is higher than the one for $\mathrm{D2}$ because, some users tend to share their Foursquare posts (check-ins) on their Twitter accounts at the same time (there is no such behavior between Google+ and Twitter).
\end{itemize}
These observations can also be generalized for other OSNs that share common behavior with the ones that we studied. We also have the following observations in terms of the attributes we used:
\begin{itemize}
	\item In both datasets, the user name attribute is the most differentiating one compared to others.
	\item We observed that the sentiment attribute does not contribute much to the profile matching in our experiments. This is possibly due to limited number of tweets, tips, and posts we processed. We expect that the contribution of this attribute might be more significant if the whole timeline is analyzed.
	\item Since the interest attribute depends on the created LDA model, as the model is created using more data, we expect more significant contribution from this attribute. Especially interest attribute would be a significant attribute for domain specific OSNs (such as health-related OSNs).
\end{itemize}
Finally, we would like to highlight that $\mathrm{Experiment~3}$ (in which we use all the attributes except for the user name) has a success rate of $38\%$ and precision and recall values up to $0.60$. This shows the threat for many OSN users even when they use completely different user names in different platforms. Therefore, in Section~\ref{sec:countermeasures}, we focus on this experiment and propose countermeasures to mitigate the risk while keeping a high profile utility for the OSN users.

\section{Countermeasures}\label{sec:countermeasures}

In this section, we propose and analyze some countermeasures against the proposed attack. More specifically, we analyze how the decrease in the utility in users' profiles (e.g., controlled obfuscation in the shared attributes of the users) affects the success of the proposed attack. Such countermeasures can be implemented in two ways: (i) the user, when he is sharing a new attribute in the target OSN, can be warned by a local software (that knows the other accounts of the user in auxiliary OSNs), or (ii) the target OSN can compute the general similarity of the user (between all the users in the auxiliary OSNs) and the user can be warned considering all his shared attributes and all the other users in the auxiliary OSNs. In the latter case, the target OSN does not really know the other accounts of the user in auxiliary OSNs, it just warns the user against a potential matching. The local software or the auxiliary OSN can also provide suggestions to the user (when the user provides new posts or attributes to the system) about the granularity of his shared material in order to prevent the profile matching. The main goal in both scenarios is to reduce the general similarity between the user in the target OSN and his profile(s) in the auxiliary OSN(s) without reducing the utility of the user in the target OSN too much.

As it is trivial for the users to pick different user names in different OSNs, and since the user name attribute does not contribute to the utility of an OSN account, in this section, we do not consider the user name attribute. Let set $U_j^T$ include all the attributes (as listed in Table~\ref{table:notation}) of a user $j$ in the target OSN $T$. Thus, $U_j^T=\{\ell_j^T,g_j^T,p_j^T,f_j^T,a_j^T,t_j^T,s_j^T\}$. We define the utility of the user $j$'s profile on the target OSN $T$ as $\Psi_j^T=\sum_{m\in{U_j^T}}c_m*\psi_m$, where $\psi_m$ and $c_m$ represent the utility of attribute $m$ and the importance (or weight) of attribute $m$ for user $j$, respectively. While the utility of an attribute can be quantified, its importance for a user is a personal preference. Based on the preference of each user, our mitigation model can compute the best privacy/utility level for each user. In the rest of this work, for the sake of simplicity, we assume that the importance of each attribute is equal.

\subsection{Attribute Utilities}

In the following, we discuss how we define the utility metric for each attribute and analyze how each attribute can be modified (i.e., how each attributes' utility is decreased) to reduce the success of the profile matching attack. Then, we analyze the privacy/utility tradeoff for the OSN users by modelling this scenario as an optimization problem. 

\subsubsection{Location}

As shown, location is one of the attributes that plays an important role in profile matching. We consider the location attribute of a user $j$ ($\ell_j^T$) as the actual location declared by the user in the target OSN (e.g., city, state, country). Thus, to reduce the utility of this attribute, we can remove the location information of target OSN users from coupled profile pairs in the test dataset or generalize it (e.g., publish the country, rather than the state or the city). Let the initial location similarity between two profiles of the user be $\hat{S}(\ell_i^A,\ell_{j}^T)$. The new (reduced) location similarity is represented as $S(\ell_i^A,\ell_{j}^T)$. Hence, location utility of the user in the target OSN due to $\ell_{j}^T$ becomes $\psi_{\ell_{j}^T}=S(\ell_i^A,\ell_{j}^T)/\hat{S}(\ell_i^A,\ell_{j}^T)$.

\subsubsection{Gender}

Gender attribute is handled differently for $\mathrm{D1}$ and $\mathrm{D2}$. In $\mathrm{D2}$ (when Google+ is the target OSN), the users explicitly reveal their gender information on the target OSN, and hence we choose whether to keep or remove this information. In $\mathrm{D1}$ (when Twitter is the target OSN), on the other hand, users do not explicitly share their gender information. Thus, we try to infer genders of the users from their user names (more specifically, from the ``name'' field of the user). To reduce the utility of the gender attribute, we modify the ``name'' field of the target OSN users. Let the initial gender similarity between two profiles of the user be $\hat{S}(g_i^A,g_j^T)$ and the new similarity be $S(g_i^A,g_j^T)$. The gender utility of the user in the target OSN is represented as $\psi_{g_j^T}=S(g_i^A,g_j^T)/\hat{S}(g_i^A,g_j^T)$. 

\subsubsection{Profile Photo}

To reduce the utility of the profile photo attribute (i.e., to reduce the photo similarity between two coupled pairs in the test dataset), we propose using alternate photos (as the profile photo of the individual) from the OSN profile of the same individual (given they exist). Thus, the photo utility of the user in the target OSN becomes $\psi_{p_j^T}=S(p_i^A,p_j^T)/\hat{S}(p_i^A,p_j^T)$ (e.g., if the profile photo remains intact, the utility remains as $1$), where $\hat{S}(p_i^A,p_j^T)$ and $S(p_i^A,p_j^T)$ represents the initial similarity and the similarity after the profile photo is changed to an alternate one, respectively. 

\subsubsection{Freetext}

To reduce the freetext utility of the coupled users in the test dataset, we gradually reduce the number of common extracted features (obtained by NER) from the information provided in both the auxiliary and the target OSNs. Thus, if the initial freetext similarity between two profiles of the user is $\hat{S}(f_i^A,f_j^T)$ and the new similarity is $S(f_i^A,f_j^T)$, the freetext utility of the user in the target OSN is represented as $\psi_{f_j^T}=S(f_i^A,f_j^T)/\hat{S}(f_i^A,f_j^T)$. In practice, this decrease in utility can be controlled directly by the user without distorting the meaning of the freetext. 

\subsubsection{Activity Pattern}

Dates and times of users' posts in the auxiliary and the target OSNs can be used to build an activity pattern for each user. As shown, this information also helps connecting the anonymized profile of a user to his unanonymized account. Especially, simultaneous posts of the user in the auxiliary and the target OSNs pose the highest risk for profile matching. To reduce the utility of a user's activity patterns, we remove some selected posts of the user from the target OSN (or post them with some latency). In particular, we remove or delay the posts that are posted close to each other in both OSNs. Thus, the activity pattern utility for a user $j$ in the target OSN is defined as $\psi_{a_j^T}=S(a_i^A,a_j^T)/\hat{S}(a_i^A,a_j^T)$, where $\hat{S}(a_i^A,a_j^T)$ and $S(a_i^A,a_j^T)$ represent the initial activity pattern similarity and the new activity pattern similarity between two profiles of the same user, respectively.

\subsubsection{Interest}

Users create posts (e.g., in terms of tips or tweets) expressing their opinions, feelings, or thoughts without being aware that this information can be used to infer their interests. We use an LDA model to infer the interests of a user from his posts in the OSNs (in Section~\ref{sec:metrics}). As shown, a user's interests in two OSNs can also play a role for connecting his two accounts to each other. We reduce the interest utility of a user by removing some of his selected posts from the target OSN. While doing so, we give the priority to the posts that increase the interest similarity between the profiles of the user in the auxiliary and target OSNs. Thus, if the initial interest similarity between two profiles of the user is $\hat{S}(t_i^A,t_j^T)$ and the new similarity is $S(t_i^A,t_j^T)$, the interest utility of the user in the target OSN is represented as $\psi_{t_j^T}=S(t_i^A,t_j^T)/\hat{S}(t_i^A,t_j^T)$.

\subsubsection{Sentiment}
As discussed, users' mood may help to connect their OSN accounts to each other. To reduce the daily sentiment utility of a user, we remove from the target OSN the posts that are posted close to each other in both OSNs. Thus, the daily sentiment utility for a user $j$ in the target OSN is defined as $\psi_{s_j^T}=S(s_i^A,s_j^T)/\hat{S}(s_i^A,s_j^T)$, where  $\hat{S}(s_i^A,s_j^T)$ and $S(s_i^A,s_j^T)$ represent the initial daily sentiment similarity and the new daily sentiment similarity between two profiles of the same user, respectively. 

\begin{table*}
	\centering
	\resizebox{\textwidth}{!}{
		\begin{tabular}{l|c|c|c|c|}
			\cline{2-5}
			& Utility & Success rate of the attack & Original attribute & Modified attribute\\
			\hline
			\multicolumn{1}{ |l| }{\multirow{2}{*}{Location}} & $1$ &  $36\%$  & \multirow{2}{*}{Hamburg, Germany} & Hamburg, Germany \\
			\multicolumn{1}{ |c| }{\multirow{2}{*}{}} & $0$ & $36\%$ & & - \\
			\hline
			\multicolumn{1}{ |l| }{\multirow{2}{*}{Gender}} & $1$  & $36\%$  & \multirow{2}{*}{Male} & Male \\
			\multicolumn{1}{ |c| }{\multirow{2}{*}{}} & $0$  & $36\%$ & & - \\
			\hline
			\multicolumn{1}{ |l| }{\multirow{3}{*}{Profile photo}} & $1$ & $36\%$  & \multirow{3}{*}{Refer to Figure~\ref{fig:utility_photo}(a)} & Refer to Figure~\ref{fig:utility_photo}(a) \\
			\multicolumn{1}{ |c| }{\multirow{3}{*}{}} & $0.5$ & $16\%$ & & Refer to Figure~\ref{fig:utility_photo}(b) \\
			\multicolumn{1}{ |c| }{\multirow{3}{*}{}} & $0$ &  $15\%$ & & - \\
			\hline
			\multicolumn{1}{ |l| }{\multirow{2}{*}{Freetext}} & $1$  & $36\%$  & \multirow{2}{*}{Sit with a pretty girl for an hour, and it seems like a minute} & Sit with a pretty girl for an hour, and it seems like a minute \\
			\multicolumn{1}{ |c| }{\multirow{2}{*}{}} & $0.5$ &  $33\%$ & & - \\
			\hline
			\multicolumn{1}{ |l| }{\multirow{2}{*}{Activity pattern}} & $1$ & $36\%$  & \multirow{2}{*}{20 posts exist} & 20 posts exist\\
			\multicolumn{1}{ |c| }{\multirow{2}{*}{}} & $0$ & $36\%$ & & 0 posts exist (all removed)\\
			\hline
			\multicolumn{1}{ |l| }{\multirow{12}{*}{Interest}} & \multirow{6}{*}{$1$} & \multirow{6}{*}{$36\%$}  & \multirow{12}{*}{\includegraphics[scale=0.30]{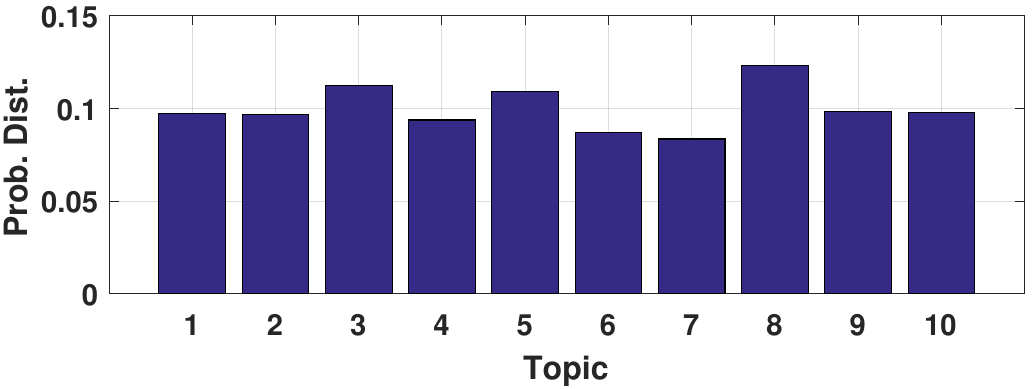}} & \multirow{6}{*}{\includegraphics[scale=0.20]{figures/topic_dist_1.pdf}} \\
			\multicolumn{1}{ |c| }{\multirow{12}{*}{}} &  &  &  &  \\
			\multicolumn{1}{ |c| }{\multirow{12}{*}{}} &  &  &  &  \\
			\multicolumn{1}{ |c| }{\multirow{12}{*}{}} &  &  &  &  \\
			\multicolumn{1}{ |c| }{\multirow{12}{*}{}} &  &  &  &  \\
			\multicolumn{1}{ |c| }{\multirow{12}{*}{}} &  &  &  &  20 posts exist\\
			\multicolumn{1}{ |c| }{\multirow{12}{*}{}} &  \multirow{6}{*}{$0.5$} & \multirow{6}{*}{$36\%$} & & \multirow{6}{*}{\includegraphics[scale=0.20]{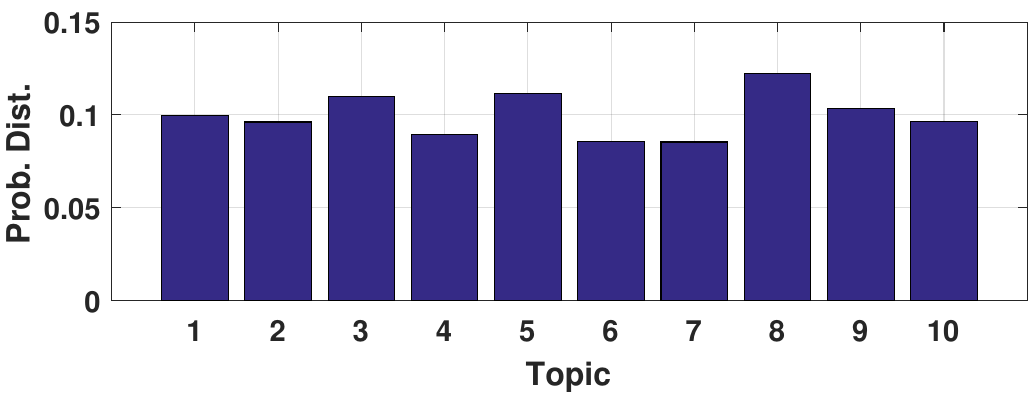}} \\
			\multicolumn{1}{ |c| }{\multirow{12}{*}{}} &  &  &  &  \\
			\multicolumn{1}{ |c| }{\multirow{12}{*}{}} &  &  &  &  \\
			\multicolumn{1}{ |c| }{\multirow{12}{*}{}} &  &  & 20 posts exist&  \\
			\multicolumn{1}{ |c| }{\multirow{12}{*}{}} &  &  &  &  \\
			\multicolumn{1}{ |c| }{\multirow{12}{*}{}} &  &  &  & 17 posts exist (3 removed)\\
			\hline
			\multicolumn{1}{ |l| }{\multirow{12}{*}{Sentiment}} & \multirow{6}{*}{$1$} & \multirow{6}{*}{$36\%$}  & \multirow{12}{*}{\includegraphics[scale=0.12]{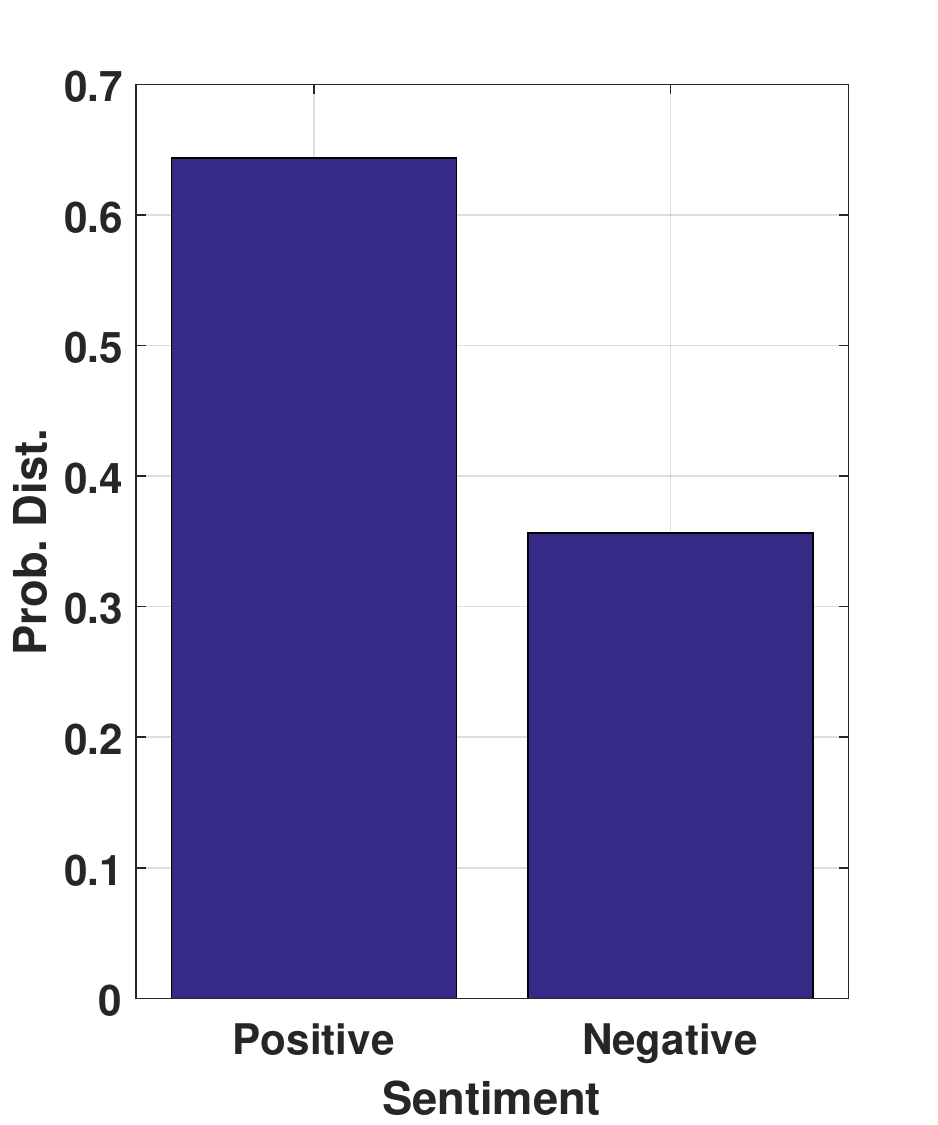}} & \multirow{6}{*}{\includegraphics[scale=0.09]{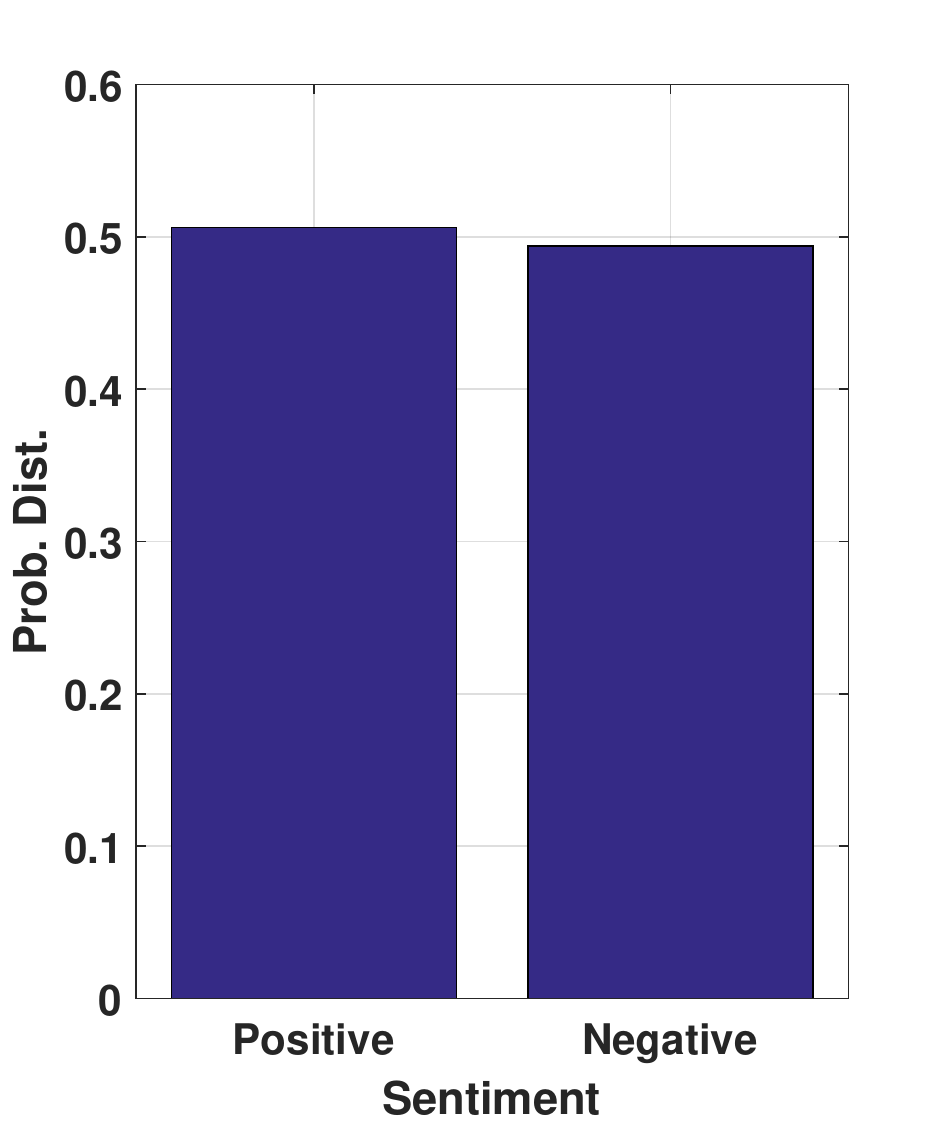}} \\
			\multicolumn{1}{ |c| }{\multirow{12}{*}{}} &  &  &  &  \\
			\multicolumn{1}{ |c| }{\multirow{12}{*}{}} &  &  &  &  \\
			\multicolumn{1}{ |c| }{\multirow{12}{*}{}} &  &  &  &  \\
			\multicolumn{1}{ |c| }{\multirow{12}{*}{}} &  &  &  &  \\
			\multicolumn{1}{ |c| }{\multirow{12}{*}{}} &  &  &  &  20 posts exist\\
			\multicolumn{1}{ |c| }{\multirow{12}{*}{}} &  \multirow{6}{*}{$0.5$} & \multirow{6}{*}{$36\%$} & & \multirow{6}{*}{\includegraphics[scale=0.09]{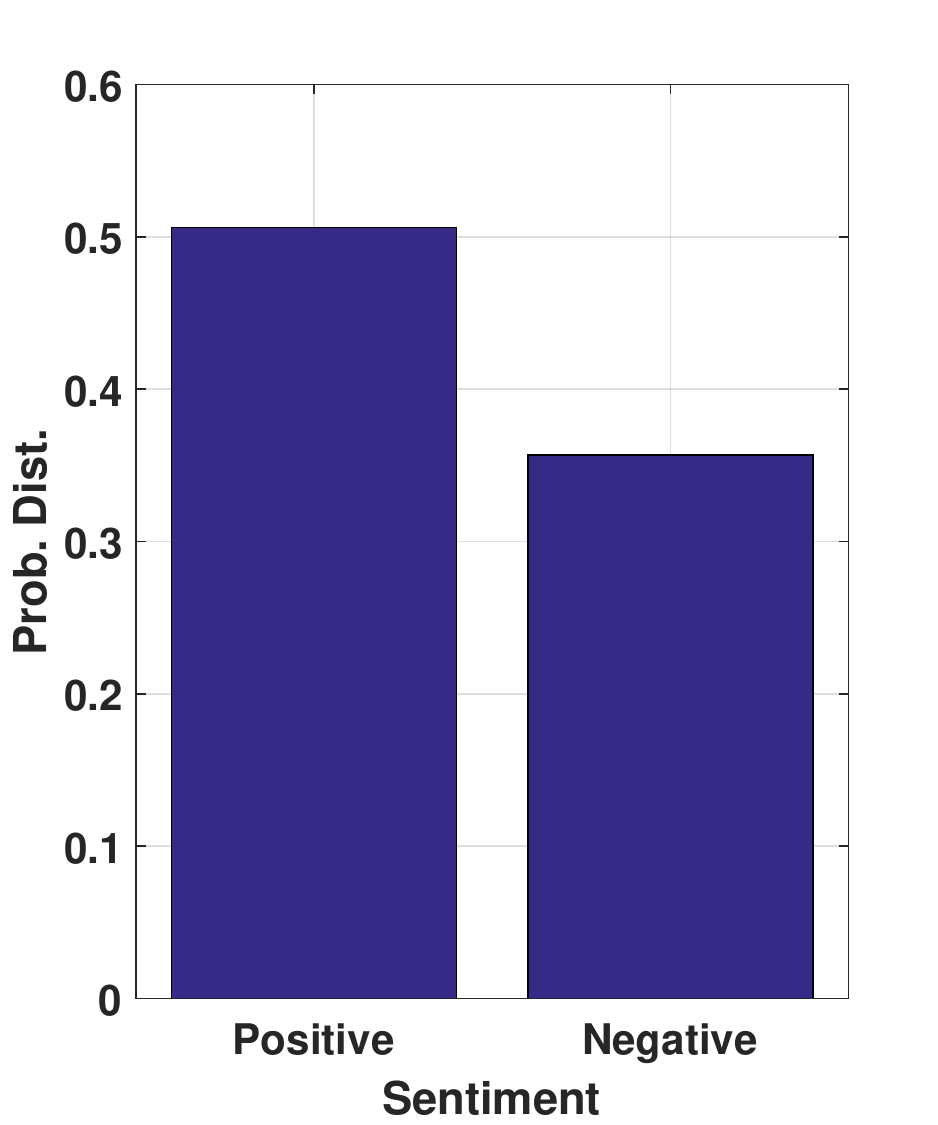}} \\
			\multicolumn{1}{ |c| }{\multirow{12}{*}{}} &  &  &  &  \\
			\multicolumn{1}{ |c| }{\multirow{12}{*}{}} &  &  &  &  \\
			\multicolumn{1}{ |c| }{\multirow{12}{*}{}} &  &  & 20 posts exist&  \\
			\multicolumn{1}{ |c| }{\multirow{12}{*}{}} &  &  &  &  \\
			\multicolumn{1}{ |c| }{\multirow{12}{*}{}} &  &  &  & 17 posts exist (3 removed)\\
			\hline
		\end{tabular}
	}
	\caption{Average individual attribute utilities computed over 500 coupled profiles with respect to the success rate of the attack. During the analysis of each individual attribute, we assume that the attacker uses all the other attributes in $\mathrm{Experiment~3}$ (in Section~\ref{sec:evaluation}) intact. ``Original attribute'' and ``modified attribute'' are shown for the target OSN profile introduced in Figure~\ref{fig:utility_profiles}(a). For the ``activity pattern'', we show the number of posts that should be removed. For the ``interest'', we show the number of removed posts and the variation in the topic distribution of the user that is obtained as a result of the created LDA model (this distribution is in particularly important as the user may receive personalized services from the target OSN based on his interests). For the ``sentiment'', we show the number of removed posts and the histogram of the user's sentiment profile.}
	\label{table:utility_individual}
\end{table*}

\subsection{Effect of Countermeasures on Privacy and Utility}

Here, we analyze the tradeoff between the privacy and utility. That is we study the optimal point at which (i) the success of the profile matching attack is minimized (i.e., target OSN users has high privacy), and (ii) the utility of the target OSN users is maximized (i.e., attributes in the profiles of target OSN users are not distorted much). Due to the space constraints, we ran our experiments only on $\mathrm{D2}$ (in which Google+ is the target OSN) by using the linear regression model. Furthermore, we only considered the global attack and we tried to match $N=500$ profiles in $\mathrm{A_e}$ to $N=500$ profiles in $\mathrm{T_e}$ (all of these profiles are coupled). Since we consider all the attributes except for the user name, we study the balance between privacy and utility under $\mathrm{Experiment~3}$ (in Section~\ref{sec:evaluation}). Under these settings, by using the original (undistorted) profiles of the users, we obtained a precision and recall of $0.6$ and $0.61$, respectively (as also shown in Table~\ref{table:coupled_attack}). Also, we observed a success rate of $36\%$. That is, we could correctly match $181$ profiles in $\mathrm{D2}$.

First, we show the effect of each attribute to the profile matching attack and show how the success rate of the attack decreases with the decrease in utility of each attribute. We gradually reduced the utility of each attribute independently for the target OSN profiles and observed how the success rate of the profile matching attack is affected from this change when only the corresponding attribute is modified and all the remaining attributes remain intact. In Table~\ref{table:utility_individual}, we report the average attribute utilities computed over these users with respect to the success rate of the attack. 

\begin{figure*}
	\centering
	\subfigure[Profile in the target OSN]{
		\includegraphics[scale=0.25]{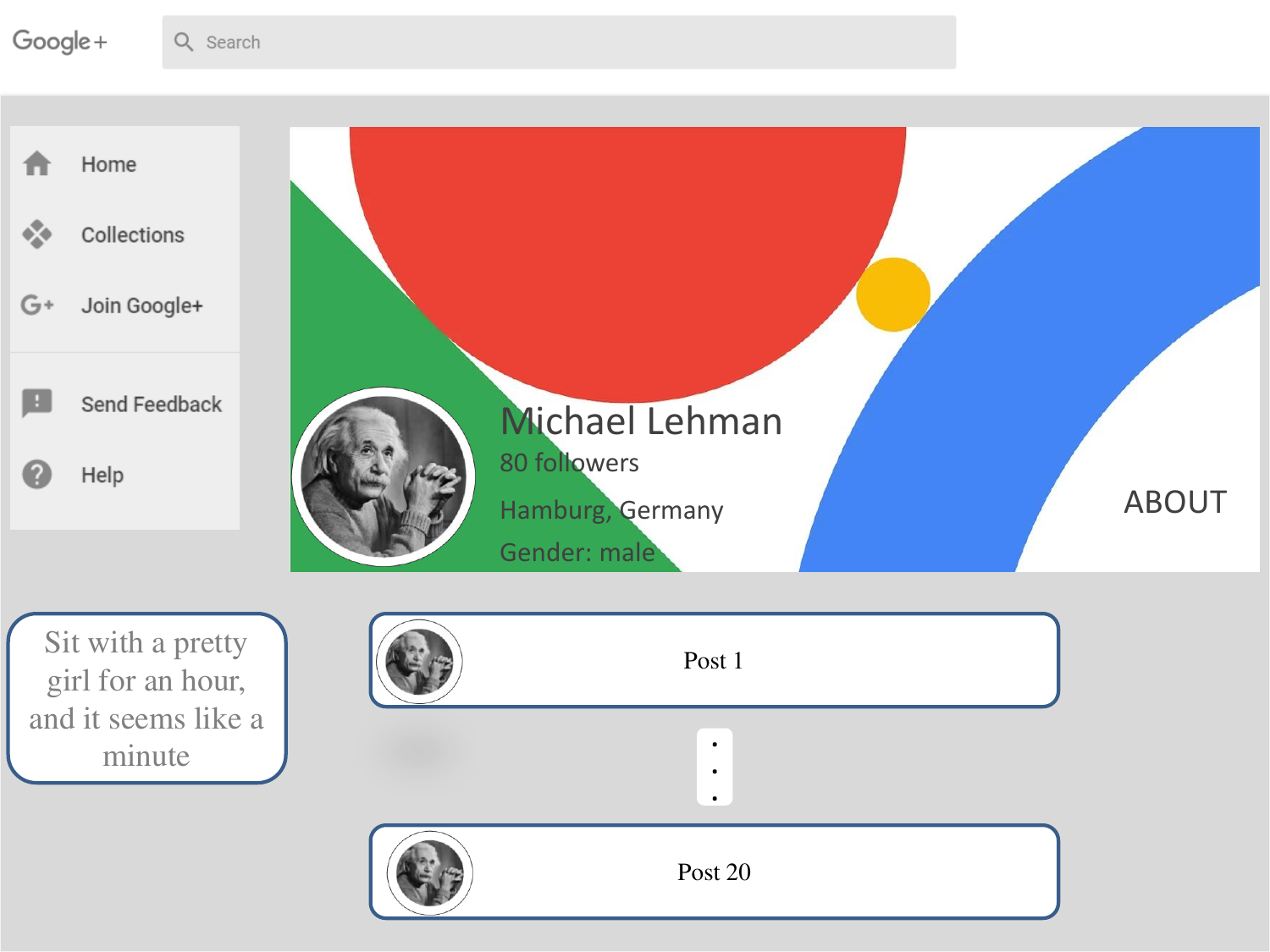}
	}
	\subfigure[Profile in the auxiliary OSN]{
		\includegraphics[scale=0.35]{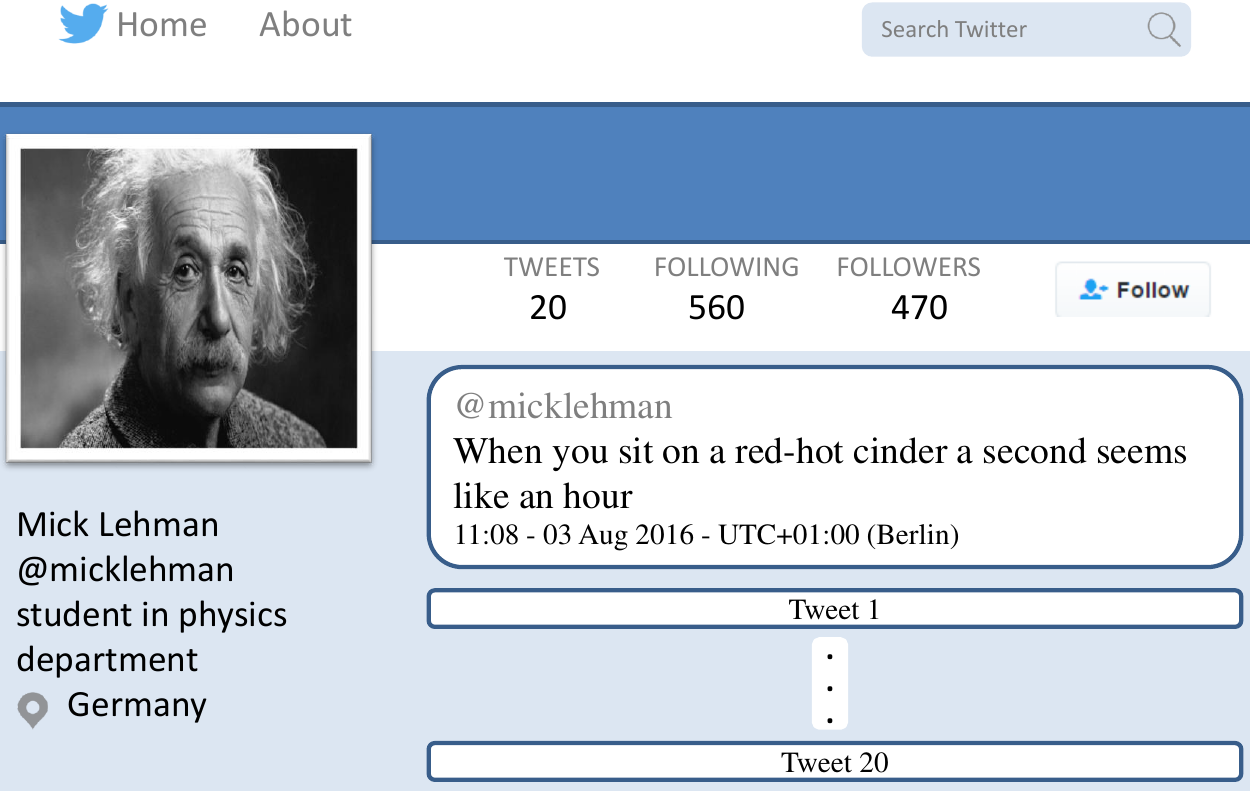}
	}
	\caption{Dummy coupled profile pair created to show the effect of decreasing utility. We also assigned 20 posts and 20 tweets to the profiles in order to evaluate the utilities for the ``activity pattern'', ``interest'', and "sentiment" attributes, however we do not show these posts and tweets due to space limitations.}
	\label{fig:utility_profiles}
\end{figure*}

For illustrative purposes, we also use a dummy coupled profile pair in Figure~\ref{fig:utility_profiles}.\footnote{Profile pair in Figure~\ref{fig:utility_profiles} does not belong to a real individual; it was created by us for illustrative purposes.}. In Table~\ref{table:utility_individual} we show how the particular target profile in Figure~\ref{fig:utility_profiles} changes with the decrease in the utility of each corresponding attribute. As the utility of an attribute decreases, the attribute should be removed partially or completely (as shown in Table~\ref{table:utility_individual}). As it was mentioned previously, profile photo is replaced by an alternate photo that provides lower photo similarity with the user profile in the auxiliary OSN. We observed that, as expected, the individual contributions (i.e., success rate) of the attributes that has lower weights (in Table~\ref{table:weights1}) to the profile matching attack are also low. Therefore, reducing the utility of such attributes, even though slightly reduces the success rate of the attack, does not have a significant impact on the overall success rate of the attack (in which all attributes are used).


\begin{figure*}[t]
	\centering
	\subfigure[Utility=1]{
		\includegraphics[height=4cm]{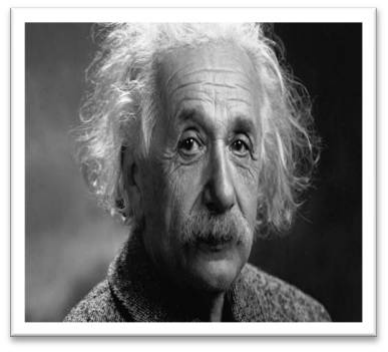}
	}
	\subfigure[Utility=0.5]{
		\includegraphics[height=4cm]{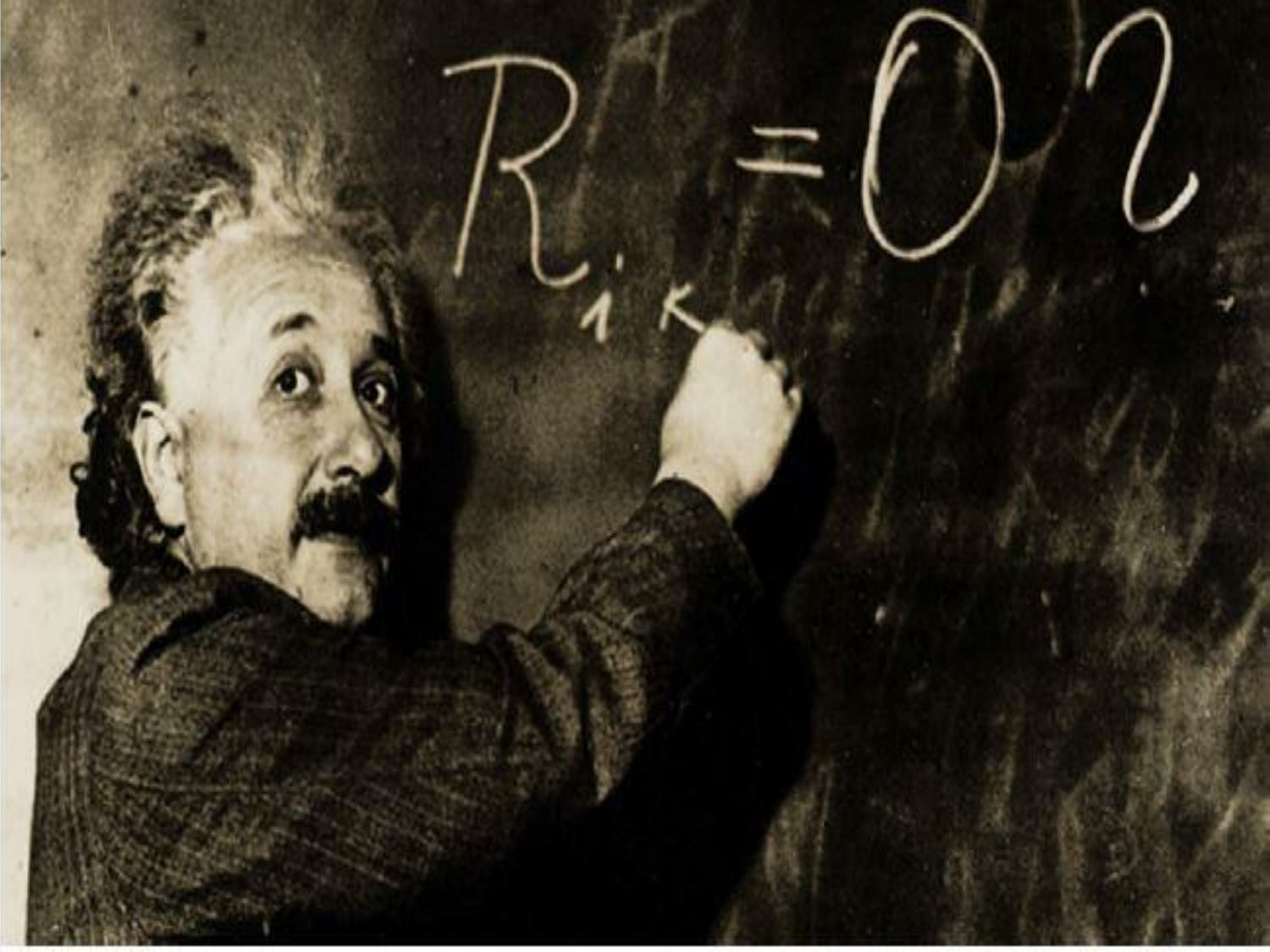}
	}
	\subfigure[Utility=0.3]{
		\includegraphics[height=4cm]{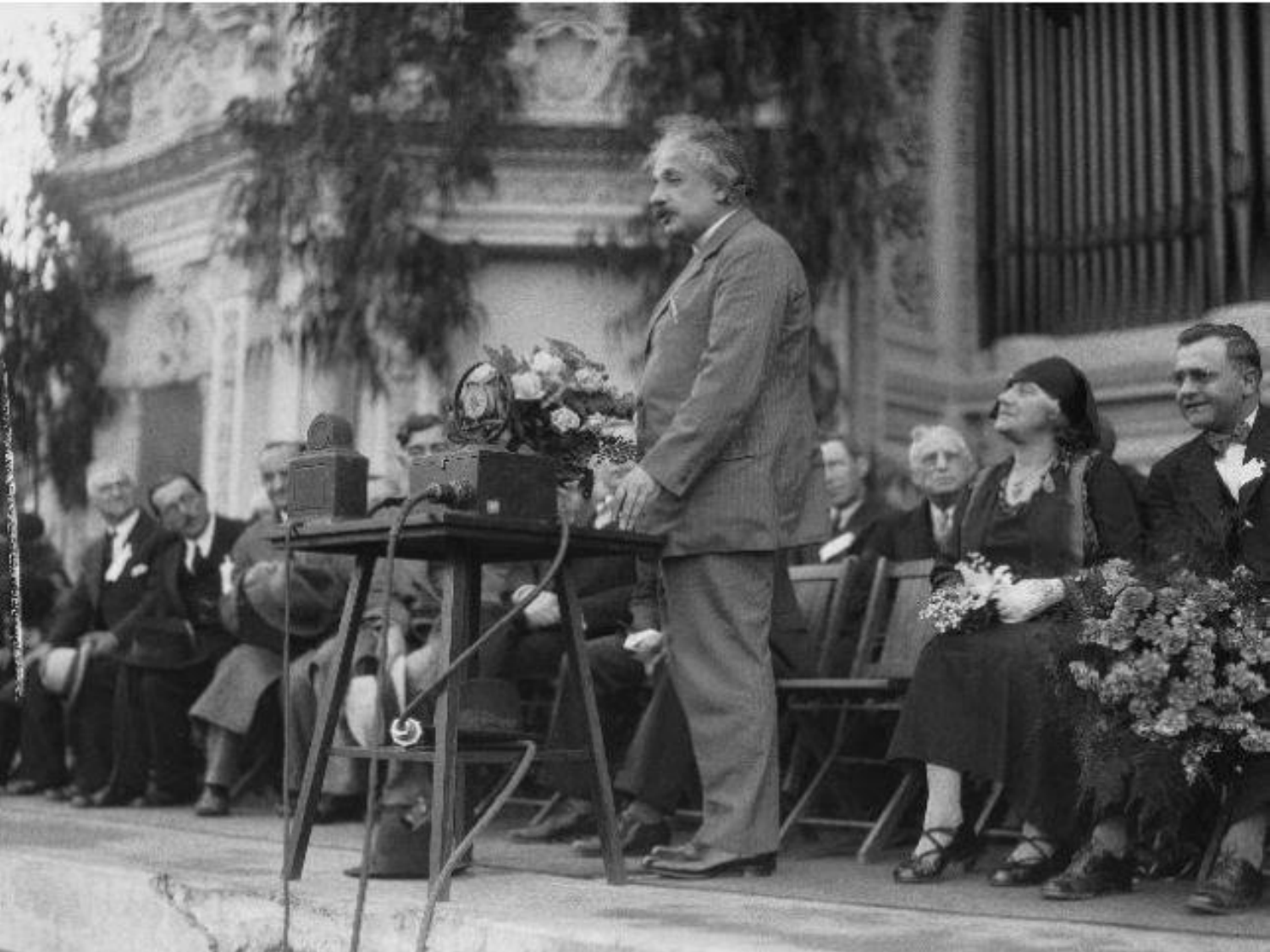}
	}
	\caption{Effect of alternative photos for the profile photo of the target OSN profile in Figure~\ref{fig:utility_profiles}(a) for three different utility values.}
	\label{fig:utility_photo}
\end{figure*}
	
Nest, we study the balance between the privacy of the target OSN users and the utility of their profiles in the target OSN. Typically, if a user has an anonymized account in the target OSN, he does not want to be deanonymized. But, at the same time, to get the best service quality from the target OSN, the user also do not want to over-distort his profile. Therefore, there is a tradeoff between his privacy and utility. We formulate this tradeoff as an optimization problem and find the optimal point that maximizes both privacy and utility of the users in the target OSN.

Let $U_i^A$ and $U_j^T$ be two coupled profiles belonging to the same individual in OSNs $A$ and $T$.  We represent the set of coupled attributes of these two profiles as $\mathrm{U}_{i,j}^{A,T}=\{(\ell_i^A,\ell_j^T), (g_i^A,g_j^T), (p_i^A,p_j^T), (f_i^A,f_j^T), \\(a_i^A,a_j^T), (t_i^A,t_j^T), (s_i^A,s_j^T)\}$. Also, let sets $W$ and $C_j$ include the weights of the attributes for profile matching\footnote{We used the weights obtained via the linear regression technique here.} and importance of each attribute (in terms of utility) for user $j$ in the target OSN. Thus, $W=\{w_{\ell},w_g,w_p,w_f,w_a,w_t, w_s\}$ and $C_j=\{c_{\ell_j^T},c_{g_j^T},c_{p_j^T},c_{f_j^T},c_{a_j^T},c_{t_j^T},c_{s_j^T}\}$. Then, for each user $j$ with the profile $U_j^T$ in the target OSN (whose profile is $U_i^A$ in the auxiliary OSN), we formulate this linear optimization problem as follows:
\begin{align}
\mathrm{maximize} \hspace{25pt} &\Psi_j^T=\sum_{m\in{U_j^T}}c_m\times\psi_m \nonumber\\
\mathrm{subject~to} \hspace{25pt} &S(U^A_i, U^T_j)\leq{\tau} \nonumber
\end{align}
where, $\psi_m$ and $c_m$ represent the utility of attribute $m$ and the importance of attribute $m$ for user $j$, respectively. For simplicity, we assume that the importance of each attribute for the user to be equal. Also, $\tau$ represents the maximum similarity value that we want to obtain between the coupled profiles. Smaller $\tau$ provides better privacy for the profiles of users in the target OSN, but smaller values of $\tau$ also causes lower profile utility. Following our previous discussion, $\Psi_j^T$ and $S(U^A_i, U^T_j)$ are computed as follows:
\begin{equation}
\Psi_j^T=\sum_{k=1}^{|\mathrm{U}_{i,j}^{A,T}|}\frac{S(\mathrm{U}_{i,j}^{A,T}(k))}{\hat{S}(\mathrm{U}_{i,j}^{A,T}(k))}\times{C_j(k)}
\end{equation}
and
\begin{equation}
S(U^A_i, U^T_j)=w_0 + \sum_{k=1}^{|\mathrm{U}_{i,j}^{A,T}|}S(\mathrm{U}_{i,j}^{A,T}(k))\times{W(k)}.
\end{equation}

The above optimization problem is similar to the bounded knapsack problem~\cite{Lagoudakis96the0-1}. To solve this problem, we use the branch-and-bound technique~\cite{shih}. In general, branch-and-bound technique finds the optimal solution and represents a good tradeoff between time and storage space. It is a systematic enumeration of all candidate solutions, where large subsets of candidate solutions are pruned by using upper bounds on the quantity being optimized. Branch-and-bound technique generally relies on two main rules: (i) the estimation of the upper bound at any node (state of assigned variables) in the search tree, and (ii) a choice criterion for the selection of a branching variable at the node selected for further partitioning.

\begin{figure}
	\centering
	\includegraphics[scale=0.30]{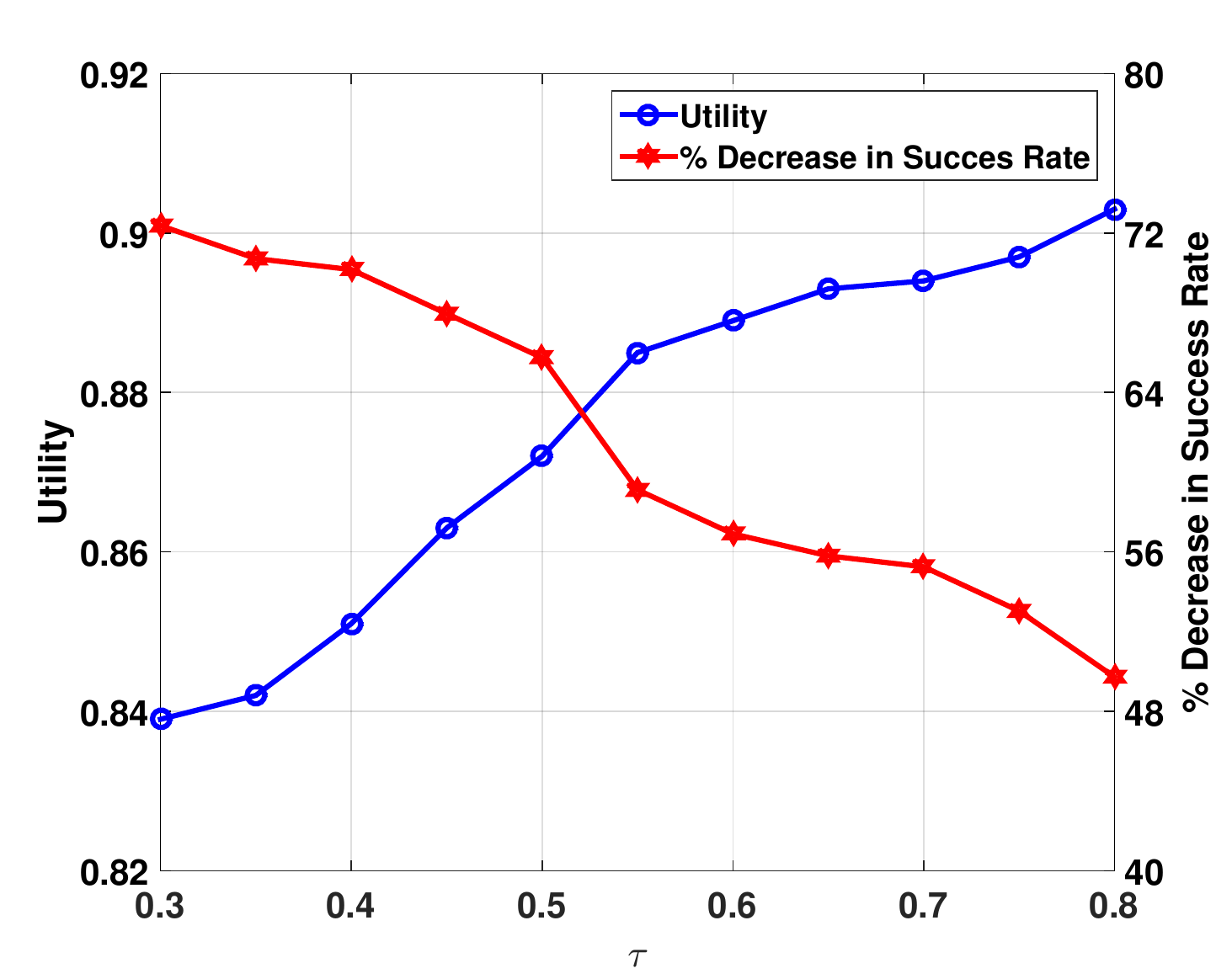}
	\caption{Solution of the optimization problem for different $\tau$ values: Variation of the average utility value for the 500 coupled profiles with respect to the decrease in success rate of the attack for different $\tau$ values.}
	\label{fig:utility_opt}
	\vspace{-15pt}
\end{figure}

We solved the above optimization problem for the coupled pairs (in the test dataset) for different values of $\tau$ and observed the decrease in the utility of the profiles. Then, we applied the Hungarian algorithm using the new similarity values and observed the decrease in the success rate of the attack. In Figure~\ref{fig:utility_opt}, we show the variation of the average utility value for the coupled  profiles with respect to the decrease in the success rate of the attack for different $\tau$ values. We observed that with a slight change in the utility of the profiles, we could reduce the success rate of the attack by more than half. Overall, we observed that the proposed optimization based countermeasure causes only slight modifications (and hence high profile utility) in the profile of the user while it significantly mitigates the profile matching attack.

\section{Related Work}\label{sec:related_work}

In the literature, most works focus on profile matching (or deanonymization) by using structural information that mainly relies on the network structure of OSN. Narayanan and Shmatikov show that statistical approaches on high-dimensional micro-data can deanonymize potentially sensitive information~\cite{narayanan2008robust}. In another work, Narayanan and Shmatikov propose a framework for analyzing privacy and anonymity in social networks and a deanonymization (DA) algorithm that is based purely on the network topology~\cite{narayan2}. Another approach by Wondracek et al. uses group membership found on social networks to identify the users~\cite{wondracek}. They show that the group membership is enough to at least reduce the number of possible candidates matching the target user. Nilizadeh et al. propose a community-level DA attack~\cite{nilizadeh} by extending the work in~\cite{narayan2}. Recently, Korula and Lattanzi propose a DA attack that starts from a set of seeds and iteratively  matches a pair of users with the most number of neighboring mapped pairs~\cite{korula}. Unlike previous attacks, Pedarsani et al. propose a seed-free DA attack~\cite{pedarsani}. It is a Bayesian based model for graph DA which uses degrees and distances to other nodes as each node's fingerprints.
Ji et al. propose a secure graph data sharing/publishing system in which they implement and evaluate graph data anonymization algorithms, data utility metrics, and modern structure-based deanonymization attacks~\cite{secgraph}. Furthermore, Ji et al. quantify deanonymizability and partial deanonymizability of real world social networks with seed information, in which a social network follows an arbitrary distribution model~\cite{quantify}. As opposed to these works, we use only publicly available unstructural data for profile matching (as mentioned, when possible, graphical structure can be used as an additional feature in our proposed model).

There are also studies that use a combination of structural and unstructural information to match user profiles. It has been shown that by using tagging information provided by users or user profile information such as user name, profile photo, description, and/or location, users in different OSNs can be linked to each other~\cite{iofciu,malhotra, nunes, vosecky}. Some works use only user names and their tags (separately or together) to link different users~\cite{iofciu}. In another work, by searching user names in Google the authors find the coupled profiles~\cite{malhotra}, however, this might not always be true. Nunes et al. apply different classifiers in the feature vectors consisting of user name, posts, and sets of friends similarities~\cite{nunes}. Furthermore, Vosecky et al. only use nick name, email, and date of birth to link the users~\cite{vosecky}. Liu et al. propose a method to match user profiles across multiple communities by using the rareness and commonness of usernames~\cite{liu2013}. They assume that usernames across social networks are consistent, which is not always the case. Zafarani et. al analyze the behaviour patterns of the users, the language used and the writing style to link users across social media sites~\cite{zafarani}. To evaluate the quality of different user attributes in profile matching, Goga et al. identify four properties: availability, consistency, non-impersonability, and discriminability~\cite{phdgoga}. They demonstrate that by using the location, activity, and writing style of a user, profiles across OSNs can be still matched.

In one of the recent works, Liu et al. propose a framework called HYDRA that uses both structural and unstructural information to match profiles across OSNs~\cite{liu2014hydra, liuarticle}. In a nutshell, the proposed framework has three steps named behavior similarity modeling, structure consistency modeling, and multi-objective optimization. However, the dataset that is used to evaluate this framework is not public, the authors only used the coupled profiles to evaluate their proposed framework, and only the obvious identifiers of the users are used for the evaluation. Peled et al. present machine learning based methods to match users profiles in different OSNs~\cite{peled}. They extract multiple features belonging to three main categories: (i) name-based features, (ii) user-information features, and (iii) social network topological-based features. By using the extracted features, different classifiers are built. The main limitation of~\cite{peled} is the fact that the coupled pairs are created only based on the user name. If the user names of the profiles match, the profile photo is used for validation. When the user names do not match, the profile pair is considered as uncoupled, which might not be always the case. In this work, we consider a wider spectrum of attributes, extensively analyze the effect of non-obvious identifiers to the profile matching attack (e.g., to quantify the risk for forum users), and propose countermeasures against the attack.

\section{Conclusion and Future Work}\label{sec:conclusion}

In this work, we have proposed a framework for profile matching in unstructured online social networks (OSNs). Our results show that using only unstructural information, users' profiles in different OSNs can be matched with high precision and success rate. We have shown how different spectrum of publicly available attributes can be utilized for such an attack. We have also shown that even a limited number of non-obvious identifiers of the users, such as activity patterns across different OSNs, interest similarities, and freetext similarities may be sufficient for the attacker in some cases. We have shown that the proposed attack significantly improves the baseline approach in terms of precision while providing comparable recall values compared to state of the art machine learning techniques. To mitigate this profile matching threat, we have also proposed possible countermeasures. In particular, we have studied the balance between the privacy and profile utility to find an optimal point that minimizes the threat and maximizes the utility the user receives from the OSN. This optimization-based technique we have proposed as a countermeasure can be widely used by the users to control their shared content on the Internet. As future work, we will work on approximate graph-matching algorithms to improve the efficiency of the proposed attack. We will also extend the work for multiple auxiliary OSNs that may have correlations with each other.



\begin{acks}
We thank Volkan K\"{u}\c{c}\"{u}k for collecting the datasets and for his help in the initial phases of this work.
\end{acks}

\bibliographystyle{ACM-Reference-Format}
\balance

\end{document}